\documentstyle[psfig,macros,xcolor,soul]{mn}
\begin{document}
\setstcolor{blue}
%

\title[Comprehensive Assessment of TBTF]
      {Comprehensive Assessment of the Too-Big-to-Fail Problem}

\author[Jiang \& van den Bosch]
       {Fangzhou Jiang\thanks{E-mail:fangzhou.jiang@yale.edu} 
        \& Frank C. van den Bosch \\
        Department of Astronomy, Yale University, New Haven, CT 06511, USA}


\date{}

\pagerange{\pageref{firstpage}--\pageref{lastpage}}
\pubyear{2014}

\maketitle

\label{firstpage}


\begin{abstract}
  We use a semi-analytical model for the substructure of dark matter
  haloes to assess the too-big-to-fail (TBTF) problem. The model
  accurately reproduces the average subhalo mass and velocity
  functions, as well as their halo-to-halo variance, in $N$-body
  simulations. We construct thousands of realizations of Milky Way
  (MW) size host haloes, allowing us to investigate the TBTF problem
  with unprecedented statistical power. We examine the dependence on
  host halo mass and cosmology, and explicitly demonstrate that a
  reliable assessment of TBTF requires large samples of hundreds of
  host haloes. We argue that previous statistics used to address TBTF
  suffer from the look-elsewhere effect and/or disregard certain
  aspects of the data on the MW satellite population. We devise a new
  statistic that is not hampered by these shortcomings, and, using
  only data on the 9 known MW satellite galaxies with $\Vmax >
  15\kms$, demonstrate that $1.4^{+3.3}_{-1.1}\%$ of MW-size host
  haloes have a subhalo population in statistical agreement with that
  of the MW. However, when using data on the MW satellite galaxies
  down to $\Vmax= 8\kms$, this MW consistent fraction plummets to $<5
  \times 10^{-4}$ (at 68\% CL). Hence, if it turns out that the
  inventory of MW satellite galaxies is complete down to $8 \kms$,
  then the maximum circular velocities of MW satellites are utterly
  inconsistent with $\Lambda$CDM predictions, unless baryonic effects
  can drastically increase the spread in $\Vmax$ values of satellite
  galaxies compared to that of their subhaloes.
\end{abstract} 




\section{Introduction} 
\label{Sec:Introduction}

The $\Lambda$+cold dark matter ($\Lambda$CDM) paradigm of structure
formation has been remarkably successful in explaining cosmic
structure covering a wide range in redshift and scale.  However, on
small scales, at low redshifts, a number of potential problems have
been identified regarding the abundance and/or structural properties
of dark matter (sub)haloes and their associated (satellite)
galaxies. In the order in which they have been introduced, these are
the `cusp-core' problem, according to which the cuspy central density
profiles of CDM haloes predicted by dark-matter-only simulations are
inconsistent with observed rotation curves (e.g., Flores \& Primack
1994; Moore 1994; Kuzio de Naray, McGaugh \& de Blok 2008;
Trachternach \etal 2008; de Blok 2010; Oh \etal 2011; but see also van
den Bosch \& Swaters 2001 and Dutton \etal 2005), the
`missing-satellite' problem, highlighting the large discrepancy
between the predicted number of CDM subhaloes per host halo and the
much smaller number of satellite galaxies detected around host
galaxies such as the Milky Way and M31 (e.g., Klypin \etal 1999; Moore
\etal 1999), and the `too big to fail' (TBTF) problem, which basically
refers to the overabundance of {\it massive, dense} subhaloes
predicted by CDM compared to the observed number of relatively {\it
  luminous} satellite galaxies of the Milky Way or the Local Group
(e.g., Boylan-Kolchin, Bullock \& Kaplinghat 2011, 2012; Martinez
2013; Garrison-Kimmel \etal 2014a, Tollerud, Boylan-Kolchin \& Bullock
2014).

In this paper we focus on the TBTF problem (hereafter simply `TBTF'),
which is generally considered the most difficult to reconcile with
$\Lambda$CDM, and which has spurred a frenzy of papers offering
possible solutions and/or advocating modifications of the standard
paradigm.  This includes suggestions to change the nature of the dark
matter from `cold' to either `warm' or `self-interacting' (e.g.,
Macc\'io \& Fontanot 2010; Vogelsberger, Zavala \& Loeb 2012; Lovell
\etal 2012; Anderhalden \etal 2013; Rocha \etal 2013; Shao \etal 2013;
Polisensky \& Ricotti 2014), relatively small changes in the
normalization, $\sigma_8$, and/or spectral index, $n_\rms$, of the
initial power spectrum (e.g., Polisensky \& Ricotti 2014), a highly
stochastic star formation efficiency for galactic subhaloes, so that a
fraction of the more massive subhaloes remain dark (e.g., Kuhlen,
Madau \& Krumholz 2013; Rodriguez-Puebla, Avila-Reese \& Drory
2013a,b), lowering the mass of the MW host halo to $\sim 10^{11.8}
h^{-1}\Msun$ (Di Cintio \etal 2011; Wang \etal 2012; Vera-Ciro \etal
2013), and enhanced tidal (impulsive) heating of satellite galaxies
due to the stellar disk of the Milky Way (Zolotov \etal 2012; Brooks
\& Zolotov 2014; Arraki \etal 2014).

Despite all this interest in TBTF, we still lack consensus regarding
both its formulation and its severity. In fact, the TBTF problem has
been formulated in a few subtly different ways.  For example, Wang
\etal (2012) express it as a problem of missing massive satellites;
high-resolution $N$-body simulations of Milky-Way (MW) size host
haloes reveal of the order of 10 subhaloes per host with a maximum
circular velocity $\Vmax \gta 25 \kms$ (e.g., Strigari \etal 2007;
Madau, Diemand \& Kuhlen 2008; Boylan-Kolchin \etal 2011). Yet, only
two MW satellite galaxies (the large and small Magellanic clouds,
hereafter LMC and SMC) are believed to have associated subhaloes that
meet this criterion. We shall refer to this as the `massive subhaloes
formulation' of TBTF. A related expression of TBTF is that the MW
reveals a gap in the $\Vmax$-distribution of its satellite galaxies.
There are no satellite galaxies known with $25\kms \lta \Vmax \lta
55\kms$, something that is not reproduced in numerical simulations
(e.g., Boylan-Kolchin \etal 2012, Cautun \etal 2014b).  We shall refer
to this as the `gap formulation' of TBTF.  Finally, in what we call
the `density formulation' of TBTF, it is argued that the (central)
{\it densities} of dark matter subhaloes are too high compared to the
central densities in satellite galaxies as inferred from their
kinematics (e.g., Boylan-Kolchin \etal 2011; Purcell \& Zentner
2012). In this paper we will compare all three formulations, and point
out how and where they differ. In particular, we will point out how
some of the solutions listed above may solve TBTF in one formulation,
but not in the other(s).

As to the severity of TBTF, the lack of consensus largely owes to poor
statistics.  On the observational side, since TBTF relates to
low-luminosity satellite galaxies, which can only be observed in the
local neighborhood, the observational `evidence' for a TBTF problem is
limited to the MW and M31. Ideally, one would like to test how common
this problem is in other host haloes, spanning a range in halo masses
and environments. On the theoretical side, the TBTF problem was
originally identified using a sample of only 6 simulated MW-size host
haloes from the Aquarius project (Springel \etal 2008). More recently,
the sample size of high resolution MW-size haloes has grown to the
order of 100, with the ELVIS (Garrison-Kimmel \etal 2014a) and
c125-2048 (Mao, Williamson \& Wechsler 2015) suites each adding of
order 50.  However, as we demonstrate in this paper, accurately
capturing the halo-to-halo variance, and characterizing the severity
of TBTF at percent level accuracy, requires of order a thousand
realizations of MW-size host haloes with subhaloes resolved down to
$\Vmax\sim10\kms$. This is immensely challenging, even for the largest
supercomputers available to date (see discussion in van den Bosch
\etal 2014).

Because of these statistical drawbacks, some authors have resorted to
semi-analytical models (Purcell \& Zenter 2012) or empirical
extrapolations of large simulations (Cautun \etal 2014a,b) to generate
large samples of well-resolved MW-size haloes.  In particular, Purcell
\& Zenter (2012; PZ12 hereafter) use a model developed by Zentner
\etal (2005) to generate thousands of MW-size haloes, and claim that
at least $\sim10$\% of MW-size haloes are free from the TBTF problem
(using the density formulation).  However, this study suffers from
three drawbacks.  First of all, the halo merger trees they use, and
which are the backbone of their analytical model, are constructed
using the method of Somerville \& Kolatt (1999), which has been shown
to be significantly and systematically biased (Zhang, Fakhouri \& Ma
2008; Jiang \& van den Bosch 2014a). Second, their recipe for
computing $\Vmax$ of subhaloes is oversimplified, resulting in
unrealistic density estimates.  And finally, they only address TBTF in
the density formulation, and it therefore remains to be seen what
their model predicts regarding the other two formulations.

In this study, we revisit the TBTF problem using a semi-analytical
model that we developed and tested in Jiang \& van den Bosch (2014b;
hereafter Paper I) and van den Bosch \& Jiang (2014; hereafter Paper
II).  The model uses accurate halo merger trees, and well-calibrated
(yet simple) recipes for the tidal stripping and disruption of
subhaloes.  We demonstrate that the model accurately reproduces the
halo-to-halo variance of subhalo properties found in numerical
simulations, and use it to gauge the severity of TBTF in all three
formulations discussed above. We examine the dependence on host halo
mass and cosmology, and explicitly demonstrate that numerical
simulation suites with of order 50 host haloes can only assess TBTF
statistics at $\sim 10$ percent accuracy. Finally, we show that all
three formulations of TBTF either are hampered by the "look-elsewhere
effect" or disregard certain aspects of the data. In order to remedy
these shortcomings, we devise a new statistic that treats all data on
equal footing, without the need to pre-identify specific $\Vmax$
scales in the data. Application of this new statistic to the existing
data on the MW satellite galaxies makes it clear that TBTF is
predominantly a statement about the $\Vmax$ distribution of MW
satellites being much broader than predicted by the $\Lambda$CDM
paradigm.

The paper is organized as follows. In \S\ref{Sec:Model} we briefly
describe the model and demonstrate that it generates subhalo
populations that are indistinguishable from those of large N-body
simulations.  \S\ref{Sec:Severity} employs thousands of model
realizations to evaluate what fraction of MW-size haloes has subhalo
populations consistent with the satellite properties of the Milky Way,
using all three formulations of the TBTF problem.  In
\S\ref{Sec:AltStat} we discuss the severity of TBTF in light of our
new statistic, and we summarize our findings in \S\ref{Sec:Summary}.


\section{Semi-analytical Model}  
\label{Sec:Model}

The basis of this work is a semi-analytical model that we developed in
Paper I.  The model is designed to generate subhalo populations for a
target host halo of any mass, in any reasonable $\Lambda$CDM
cosmology.  This section describes the model, and shows that it yields
subhalo statistics in excellent agreement with state-of-the-art
numerical simulations.  However, we start with a brief introduction of
halo basics, outlining a number of definitions and notations.

\subsection{Halo Basics}
\label{Sec:Basics}

Throughout this paper, dark matter haloes at redshift $z$ are defined
as spherical systems with a virial radius $\Rvir$ inside of which the
average density is equal to $\Delta_{\rm vir}(z)\rho_{\rm crit}(z)$.
Here $\rho_{\rm crit}(z)=3H^2(z)/8\pi G$ is the critical density, and
$\Delta_{\rm vir}(z)$ is given by
\begin{equation} \label{Eq:VirialOverdensity}
\Delta_{\rm vir}(z)=18\pi^2+82x-39x^2
\end{equation}
with $x = \Omega_\rmm(z) - 1$ (Bryan \& Norman 1998).  Haloes whose
center falls within the virial radius of another halo are called
subhaloes, while haloes that are not subhaloes are called host
haloes. The (virial) mass of a host halo is defined as the mass within
the virial radius $\Rvir$ and indicated by $M$. The mass of a subhalo
is defined as the fraction of its mass at accretion (i.e., when it
transits from being a host halo to a subhalo) that remains bound, and
is indicated by $m$ (see Paper~II for an in-depth discussion of
subtleties associated with mass definitions).

Throughout we assume that host haloes have a NFW density profile
(Navarro, Frenk \& White 1997), characterized by a concentration
parameter $c = \Rvir/\Rs$, with $\Rs$ the NFW scale radius, and a
maximum circular velocity of
\begin{equation} \label{Eq:Vmax}
\Vmax= 0.465 \Vvir \sqrt{\frac{c}{\ln(1+c)-c/(1+c)}},
\end{equation}
Here
\begin{eqnarray} \label{Eq:Vvir}
\lefteqn{ \Vvir = \sqrt{G M \over \Rvir} = 
159.43\kms\left(\frac{M}{10^{12}\Msunh}\right)^{1/3}} \nonumber \\
& & \times \left[\frac{H(z)}{H_0}\right]^{1/3}\left[\frac{\Delta_{\rm vir}(z)}{178}\right]^{1/6}
\end{eqnarray}
is the virial velocity, and the radius, $r_{\rm max}$, at which the
circular velocity reaches its maximum value, is given by 
\begin{equation} \label{Eq:Rmax}
\Rmax = 2.163 \, \Rs\,.
\end{equation}

\subsection{Model Description}  
\label{Sec:ModelOverview}

The semi-analytical model that we use to construct realizations of
dark matter subhalo populations is a strongly improved and modified
version of the model introduced in van den Bosch, Tormen \& Giocoli
(2005; hereafter vdB05). It treats subhalo mass stripping in an
orbit-averaged sense, allowing the construction of thousands of
realizations in a manner of minutes. This section gives a brief
overview of the model ingredients. A more detailed description can be
found in Paper~I.

The backbone of our model, as for any other semi-analytical model for
the substructure of dark matter haloes (e.g., Taylor \& Babul 2001,
2004, 2005a,b; Benson \etal 2002; Taffoni \etal 2003; Oguri \& Lee
2004; Zentner \& Bullock 2003; Pe\~{n}arrubia \& Benson 2005; Zentner
\etal 2005; vdB05; Gan \etal 2010; Yang \etal 2011; Purcell \& Zentner
2012), is halo merger trees. These describe the hierarchical mass
assembly of dark matter haloes, and therefore yield the masses,
$\macc$, and redshifts, $\zacc$, at which the dark matter
subhaloes are accreted into their hosts. We also use the merger trees
to compute the concentration parameter, $\cacc$, of the subhalo
at accretion. For this we use the fact that halo concentration is
tightly correlated with its assembly history (e.g., Wechsler \etal 2002;
Ludlow \etal 2013). In particular, we compute $\cacc$ using
the model of Zhao \etal (2009), according to which
\begin{equation} \label{Eq:Concentration}
c(t) = 4.0\left(1+\left[\frac{t}{3.75t_{0.04}}\right]^{8.4}\right)^{1/8}
\end{equation}
Here $t_{0.04}$ is the time at which the main progenitor of the halo
in question assembled 4 percent of its mass at time $t$, which we
compute from our merger tree by tracing the assembly history of the
subhalo back in time. Using eqs.~(\ref{Eq:Vmax})--(\ref{Eq:Rmax}),
combined with $\macc$, $\zacc$ and $\cacc$, we also compute $\Vmax$
and $\Rmax$ of the subhalo at accretion, which we indicate by $\Vacc$
and $\Racc$, respectively.

After accretion, a subhalo orbits its host halo, subject to tidal
stripping, impulsive heating and dynamical friction. Following vdB05
we model the tidal stripping in an orbit-averaged sense, where the
average is taken over all orbital energies, eccentricities and phases.
Using a simple toy model, in which it is assumed that over a radial
orbital period the subhalo is stripped of all mass outside of the
subhalo's tidal radius at the orbit's pericentric distance from the
center of the host halo (see Paper~I for details), one predicts
an orbit-averaged mass loss rate that is well described by
\begin{equation} \label{Eq:MassLoss}
{\rmd m \over \rmd t} = \calA 
\frac{m}{\tau_{\rm dyn}}\left(\frac{m}{M}\right)^\zeta,
\end{equation}
with $m$, $M$, and $\tau_{\rm dyn}$ the instantaneous subhalo mass,
host halo mass, and host halo's dynamical time, respectively.  The
same functional form was also used by Giocoli, Tormen \& van den Bosch
(2008) to fit the orbit averaged mass loss rates of dark matter
subhaloes in numerical $N$-body simulations, which yielded $\calA =
1.54^{+0.52}_{-0.31}$ and $\zeta = 0.07\pm 0.03$. Our toy model
predicts that $\zeta \simeq 0.04$ and that $\calA$ follows a
log-normal distribution with median $\bar{\calA} \simeq 0.81$ and
scatter $\sigma_{\log\calA} \simeq 0.17$. This scatter in $\bar{A}$ is
due to the variance in orbital properties (energy and angular
momentum) and halo concentrations (of both the host and the
subhalo). Given the uncertainties in the parameters derived from the
simulations, and the oversimplifications of the toy model, we treat
$\calA$ and $\zeta$ as free parameters. As detailed in Paper~I, we
tune $\bar{\calA}$ and $\zeta$ so as to reproduce the subhalo mass
functions in the Bolshoi simulation (see \S\ref{Sec:Simulations}
below). This results in $\bar{\calA}=0.86$ and $\zeta = 0.07$, close
to the values suggested by the toy model, and in good agreement with
the simulation results of Giocoli \etal (2008). The scatter we keep
fixed at $\sigma_{\log\calA} = 0.17$. For each individual subhalo in
our model, we randomly draw a value for $\calA$ from the log-normal,
and evolve its mass using Eq.~(\ref{Eq:MassLoss}).

During its pericentric passage, a subhalo experiences impulsive
heating which increases the kinetic energy of its constituent
particles. Depending on the amount of energy transferred, the subhalo
either expands (`puffs up') while re-establishing virial equilibrium,
or is tidally disrupted (typically if the kinetic energy injected
exceeds the gravitational binding energy of the subhalo). In our model,
a subhalo is disrupted (i.e., removed from the inventory) if its mass
$m(t)$ drops below a critical mass
\begin{equation} \label{Eq:Disruption}
m_{\rm dis} \equiv m_{\rm acc}(<f_{\rm dis} r_{\rm s,acc}),
\end{equation}
with $m_{\rm acc}(<r)$ the mass enclosed within radius $r$ at
accretion, and $r_{\rm s,acc}$ the NFW scale radius of the subhalo at
accretion.  The dependence on $r_{\rm s,acc}$ assures that subhaloes
that are more concentrated at accretion are more resistant to
disruption.  Using idealized $N$-body simulations, Hayashi \etal
(2003) found that $f_{\rm dis} \simeq 2$, while both Taylor \& Babul
(2004) and Zentner \etal (2005) used Eq.~(\ref{Eq:Disruption}) to
model tidal disruption in their semi-analytical models, but with
wildly differing values of $f_{\rm dis} = 0.1$ and $f_{\rm dis} =
1.0$, respectively. As detailed in Paper~I, tidal disruption in
numerical $N$-body simulation occurs for values of $f_{\rm dis}$ that
follow a broad (roughly log-normal) distribution with a median of
$f_{\rm dis} \sim 1.5$ and standard deviation $\sigma_{\log(f_{\rm
    dis})} \sim 0.55$. Based on these findings, for each subhalo we
draw a value of $f_{\rm dis}$ from a similar log-normal distribution,
and disrupt the subhalo whenever its mass drops below $m_{\rm
  dis}$. By calibrating the model to reproduce the distribution of
retained mass fractions, $m/m_{\rm acc}$, of the surviving subhaloes in
the Bolshoi simulation, we find that we need to introduce a weak mass
dependence in $f_{\rm dis}$, and we end up modeling the $f_{\rm dis}$
distribution as a log-normal with median
\begin{equation}\label{fdisrupt}
\bar{f}_{\rm dis} = 1.5 \left[ 1 + 0.8 \, \left\vert 
\log\left({m_{\rm acc} \over M_{\rm acc}}\right) \right\vert^{-3}\right]
\end{equation}
and standard deviation $\sigma_{\log(f_{\rm dis})}=0.55$.

The model for mass stripping and tidal disruption described above
regulates the mass evolution of the subhaloes. However, in order to
address TBTF, we also need to model the structure (i.e., density
distribution) of the individual subhaloes. In particular, we need to
be able to predict $\Vmax$ and $\Rmax$. Based on the idealized
simulations of Pe\~{n}arrubia \etal (2008, 2010), we model $\Vmax$
and $\Rmax$ of the subhaloes using
\begin{equation} \label{Eq:P10}
{\Vmax \over \Vacc} = 1.32 {x^{0.3} \over (1 + x)^{0.4}}\,, 
\;\;\;\;\;\;\;\;\;\;
{\Rmax \over \Racc} = 0.81 {x^{0.4} \over (1 + x)^{-0.3}}\,,
\end{equation}
with $x = m/\macc$; i.e., the evolution of $\Vmax$ and $\Rmax$ of
subhaloes depends only on the {\it amount} of mass loss, but not on
{\it how} that mass was lost (see also Hayashi \etal 2003). As shown
in Paper~I, Eq.~(\ref{Eq:P10}) is in good agreement with results from
various numerical simulations.

\subsection{N-body Simulations}
\label{Sec:Simulations}

In this paper we use several numerical $N$-body simulations for
comparison with our model predictions. The first is the Bolshoi
simulation (Klypin, Trujillo-Gomez \& Primack 2011), which follows the
evolution of $2048^3$ dark matter particles in a box of size
$250\mpch$ using the Adaptive Refinement Tree (ART) code (Kravtsov,
Klypin \& Khokhlov 1997) in a flat $\Lambda$CDM cosmology with
parameters $(\Omega_{\rmm,0},\Omega_{\Lambda,0},
\Omega_{\rmb,0},h,\sigma_8, n_\rms)=(0.27,0.73,0.047,0.7,0.82,0.95)$
(hereafter `Bolshoi cosmology'). With a particle mass $m_\rmp = 1.4
\times 10^8 \Msunh$, Bolshoi resolves haloes down to $\sim 10^{10}
\Msunh$ (corresponding to $\Vmax \sim 50\kms$).  We use the publicly
available halo
catalogs\footnote{http://www.slac.stanford.edu/$\sim$behroozi/Bolshoi\_Catalogs/}
obtained using the phase-space halo finder \Rockstar (Behroozi \etal
2013a,b).  \Rockstar haloes are defined as spheres with an average
density of $\Delta_{\rm vir}\rho_{\rm crit}$, in line with the halo
definition used throughout this paper.

In addition to the Bolshoi simulation, we also compare our results to
the ELVIS suite of zoom-in simulations of 48 MW-size dark matter
haloes (Garrison-Kimmel \etal 2014a). These cover the mass range
$11.85 < \log[M/(\Msunh)] < 12.31$, comparable to the range of masses
quoted for the Milky Way in the literature. The haloes have been
selected from medium-resolution cosmological volumes of box size
$50\mpch$ and re-simulated with progressively higher resolution up to
$m_\rmp = 1.35 \times 10^5\Msunh$. The ELVIS suite adopts a flat
$\Lambda$CDM cosmology with $(\Omega_{\rmm,0},\Omega_{\Lambda,0},
\Omega_{\rmb,0},h,\sigma_8, n_\rms) =
(0.266,0.734,0.045,0.7,0.801,0.963)$ (hereafter `WMAP7 cosmology),
which are the parameters that best fit the 7-year data release of the
Wilkinson Microwave Anisotropy Probe (Larson \etal 2011). The halo
catalogs of the ELVIS suite are publicly
available\footnote{http://localgroup.ps.uci.edu/elvis/data.html}, and
contain subhaloes (identified with \Rockstar, and using the same halo
definition as for Bolshoi), down to the resolution limit of $V_{\rm
  max} = 8\kms$.

\subsection{Comparison with Simulations}  
\label{Sec:Comparison}

In this section, we first summarize the results of Paper I \& II by
showing that the model accurately matches the {\it average} subhalo
mass and $\Vmax$ functions extracted from $N$-body simulations, and
then demonstrate that the same model, without any modifications, also
accurately reproduces the halo-to-halo variance of subhalo statistics.
\begin{figure*}
\centerline{\psfig{figure=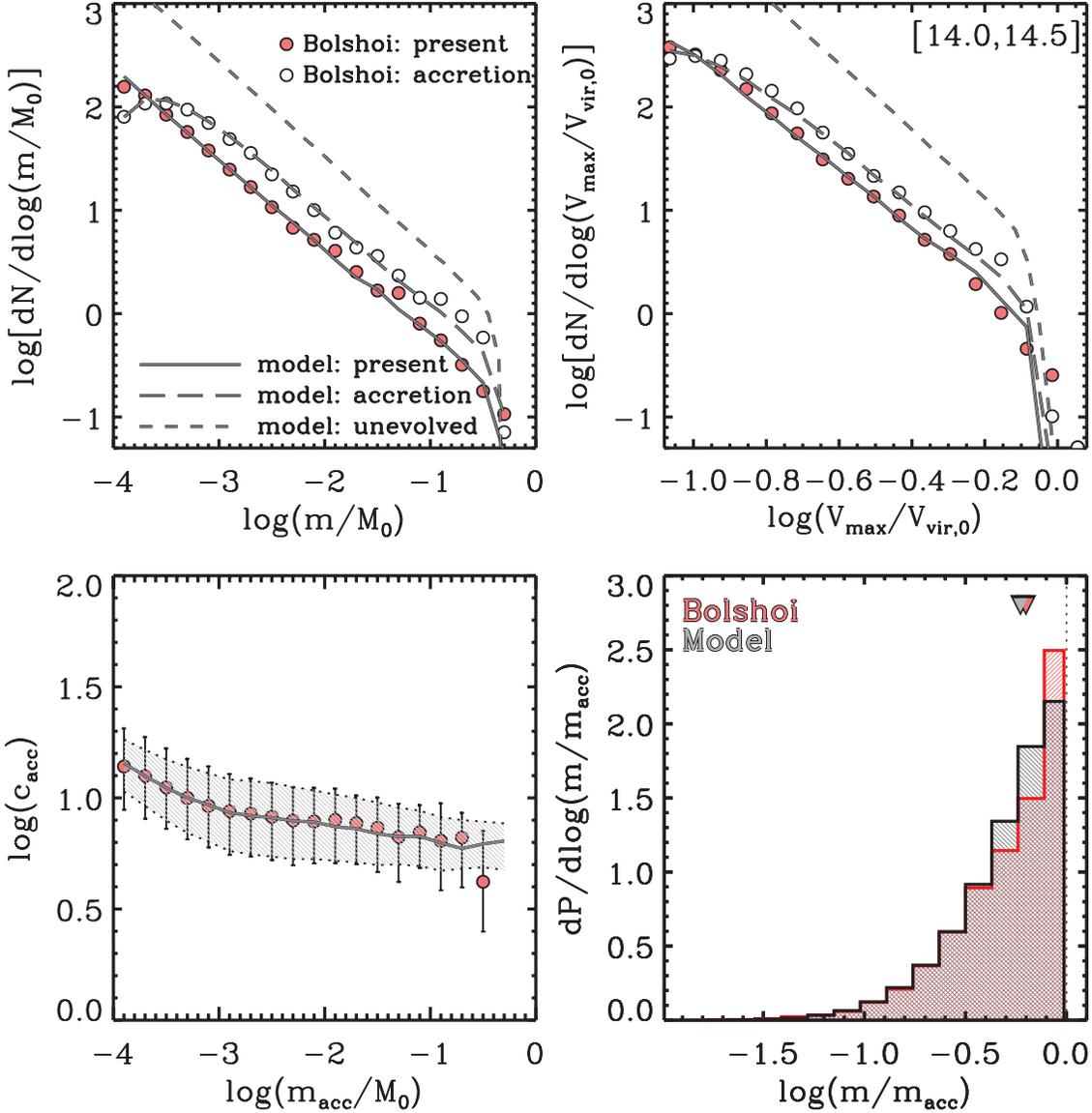,width=0.85\hdsize}}
\caption{The upper panels compare the average subhalo mass (left) and
  velocity (right) functions of the Bolshoi simulation (circles) and
  our model (lines).  The line- and symbol- styles differentiate the
  results for the present-day surviving subhaloes (`present'), the
  surviving subhaloes at accretion (`accretion'), and all subhaloes
  ever accreted (`unevolved').  The lower, left-hand panel plots the
  concentrations of the surviving subhaloes at accretion, $c_{\rm
    acc}$, as a function of their masses at accretion (normalized to
  the present-day host halo mass).  Symbols (Bolshoi) and solid curve
  (model) represent the median relations, while error bars and hatched
  region indicate the 68\% confidence intervals. Finally, the lower,
  right-hand panel shows the distributions of the retained-mass
  fractions of the surviving subhaloes, with downward triangles
  indicating the medians.}
\label{Fig:Calibration}
\end{figure*}

The upper panels of Fig.~\ref{Fig:Calibration} compare the model
predictions for the average subhalo mass function, $\rmd
N/\rmd\log(m/M_0)$, and the average subhalo velocity function, $\rmd
N/\rmd\log(\Vmax/V_{\rm vir,0})$ (solid and long-dashed curves), with
the results from the Bolshoi simulation (filled and open circles). The
latter are averaged over a total of 281 host haloes with mass $M_0 =
10^{14.25\pm0.25}\Msunh$, while the model predictions are obtained
averaging over 2000 realizations of host haloes with mass $M_0 =
10^{14.25}\Msunh$, adopting the Bolshoi cosmology.  Solid lines and
filled circles indicate the abundances of the surviving subhaloes as a
function of their {\it present-day} mass and $\Vmax$, while dashed
lines and open circles show the abundances of the same subhaloes, but
as function of their mass and $\Vmax$ {\it at accretion}. In both
cases, the model predictions are in excellent agreement with the
Bolshoi results. Note that the abundance functions {\it at accretion}
are not to be confused with the {\it unevolved} subhalo mass and/or
$\Vmax$ functions, and which are indicated by the short-dashed curves.
The latter include {\it all} subhaloes that have {\it ever} been
accreted onto the main progenitor of the host halo, and includes those
subhaloes that have been disrupted since. Note that the unevolved
subhalo mass and velocity functions are substantially higher than
those for the surviving population, indicating that subhalo disruption
is extremely efficient and important.

The lower, left-hand panel of Fig.~\ref{Fig:Calibration} plots the
concentrations at accretion of the present-day, surviving subhaloes as
a function of their mass at accretion.  The model prediction is again
in excellent agreement with Bolshoi, indicating that our model
accurately reproduces the structural properties of subhaloes at infall.
Finally, the lower, right-hand panel of Fig.~\ref{Fig:Calibration}
compares the distributions of the retained mass fraction, $m/m_{\rm
  acc}$, of the present-day, surviving subhaloes.  Note that very few
subhaloes lose more than 90\% of their initial mass, which is another
manifestation of efficient subhalo disruption.  Here the good
agreement between the model and Bolshoi lays the foundation for the
accurate model prediction of the parameters that are important for
TBTF, $\Vmax$ and $\Rmax$, whose evolution is purely a function of
$m/m_{\rm acc}$ (see Eq.~[\ref{Eq:P10}])

The set of comparisons shown in Fig.~\ref{Fig:Calibration} are the key
diagnostics that we used in Paper~I to calibrate our model. Note,
though, that the comparison is for host haloes with $M_0 \sim
10^{14.25}\Msunh$, far from the mass scale of interest for the TBTF
problem. The reason for showing this mass scale is that, in the
Bolshoi simulation, it probes a large dynamic range in subhalo mass
(down to $m = 10^{-4} M_0$) and a sufficiently large sample size
within the simulation box to be able to compute a reliable average
mass function. As we have demonstrated in detail in Papers~I and II,
the same model is equally successful at other mass scales, as far as
they have been probed by simulations.
\begin{figure*}
\centerline{\psfig{figure=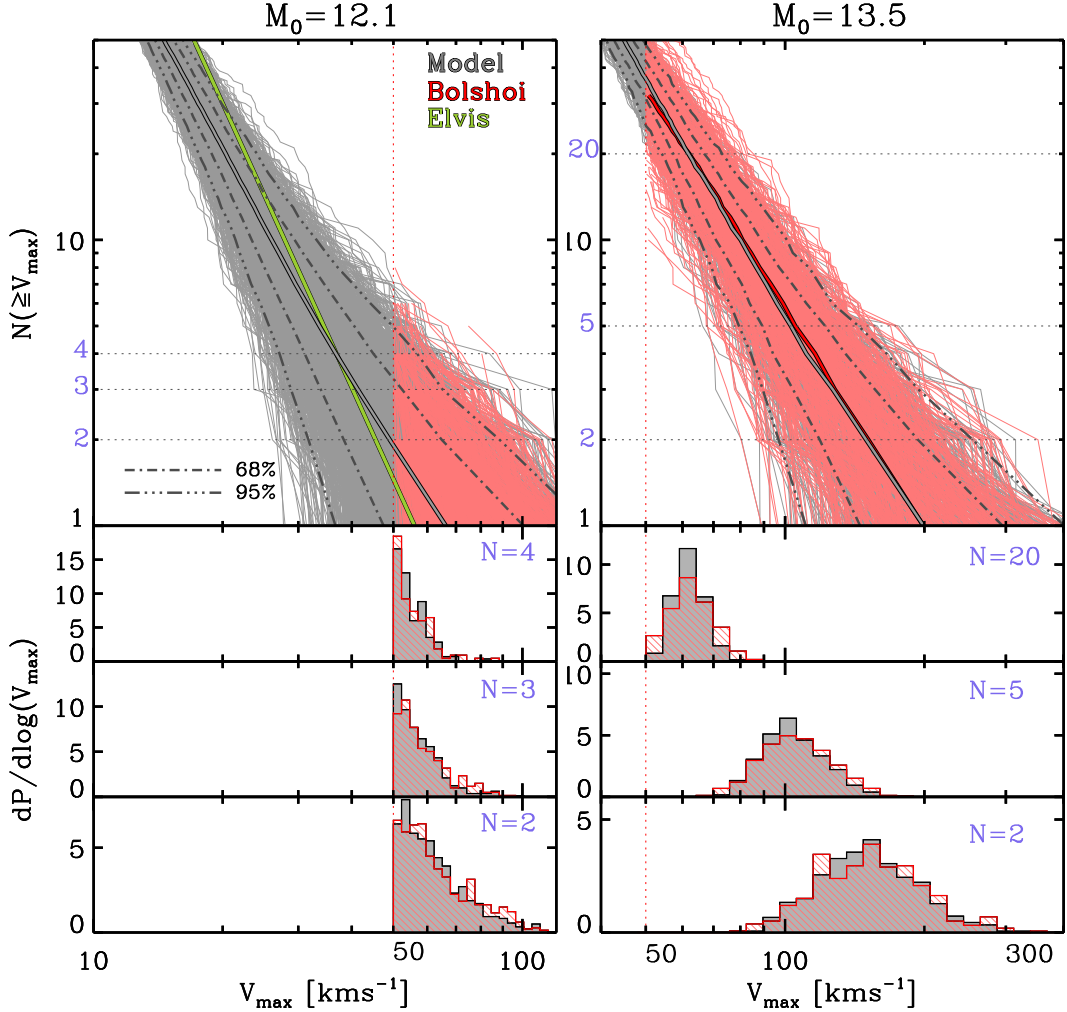,width=0.8\hdsize}}
\caption{Cumulative subhalo velocity functions for individual host
  haloes.  {\it Upper, left-hand panel}: 2000 model realizations of
  haloes with $M_0=10^{12.10}\Msunh$ (grey) and 1986 Bolshoi haloes
  with $M_0=10^{12.10\pm 0.01}\Msunh$ (red).  {\it Upper, right-hand
    panel}: 500 model realizations of haloes with
  $M_0=10^{13.50}\Msunh$ (grey) and 441 Bolshoi haloes with
  $M_0=10^{13.50\pm 0.05}\Msunh$ (red).  The vertical, dotted lines
  mark the Bolshoi resolution limit at $V_{\rm max}=50\kms$.  The
  thick, grey line indicates the model's median $\Vmax$ at given
  $N(\geq \Vmax)$, with the dash-dotted lines bracketing the 68\% and
  95\% intervals, as indicated.  The thick, red curve (upper,
  right-hand) is the median relation for the Bolshoi simulation.  The
  thick, green curve (upper, left-hand) is the best-fit median
  relation for the ELVIS simulation, as given by Garrison-Kimmel \etal
  (2014a).  {\it Bottom panels:} $\Vmax$ distribution of the $N^{\rm
    th}$ subhalo rank-ordered in $V_{\rm max}$.  }
\label{Fig:Scatter}
\end{figure*}

Fig.~\ref{Fig:Scatter} plots the {\it individual}, cumulative subhalo
velocity functions, $N(\geq \Vmax)$, for 1986 host haloes with mass
$M_0 = 10^{12.10\pm0.01}\Msunh$ (left-hand panel) and 441 host haloes
with mass $M_0 = 10^{13.50\pm0.05}\Msunh$ (right-hand panel) in the
Bolshoi simulation, and compares them with 2000 model realizations of
halo mass $M_0 = 10^{12.10}\Msunh$, and 500 model realizations of halo
mass $M_0 = 10^{13.50}\Msunh$, respectively. The Bolshoi results are
plotted down to $\Vmax = 50\kms$, which corresponds to roughly 250
particles, the minimum number of particles required to resolve haloes
well enough for a reliable estimate of $\Vmax$ (see Paper II).  For
the model realizations, we trace subhaloes down to a mass of
$10^{-5}M_0$. In the case of host haloes with $M_0 = 10^{12.10}\Msunh$
this roughly translates to $\Vmax = 10\kms$, which is more than
sufficient for a detailed assessment of TBTF.

The model predictions manifest excellent agreement with the Bolshoi
results for both mass scales. In the case of $M_0 \sim
10^{13.50}\Msunh$, for which the median in Bolshoi can be measured
over an appreciable range in $\Vmax$, the ensemble average of the
model predictions (indicated by the thick gray line) agrees almost
perfectly with that of the Bolshoi simulation data (solid red line).
In terms of the halo-to-halo variance, the model and Bolshoi results
are almost indistinguishable down to the resolution limit of the
simulation.  This is nicely illustrated in the bottom panels of
Fig.~\ref{Fig:Scatter}, which show the $\Vmax$ distributions at
several values of $N(\geq \Vmax)$, as indicated.

The average, cumulative subhalo velocity functions are usually
described by a simple power law, $N(\geq \Vmax) \propto
\Vmax^{-\alpha}$ (at least for $N \gta 2$).  The Bolshoi simulation,
against which our model is calibrated, yields a slope of 
$\alpha\approx 2.9$ with, as far as we can tell given the limited mass
resolution, no dependence on host halo mass (see Papers I \& II).
This is in excellent agreement with our model predictions, which also
yields $\alpha\approx 2.9$
without any significant dependence on host halo
mass.  The ELVIS suite, however, seems to predict a slope that is
significantly steeper, with $\alpha = 3.3$ (Garrison-Kimmel \etal
2014a). It is unclear what the cause is of this discrepancy. We
emphasize that most other simulations all suggest that $\alpha \lta 3$
(e.g., Klypin \etal 1999; Reed \etal 2005; Madau \etal 2008; Diemand
\etal 2008; Wu \etal 2013). The only simulations for which similarly
large values for $\alpha$ have been reported are the Millennium-II
simulation (Boylan-Kolchin \etal 2009) and the suite of Aquarius
simulations (Springel \etal 2008). Both of these were analyzed with
the halo finder \Subfind (Springel, White \& Hernquist 2001), which
has been shown to give much steeper subhalo mass and velocity
functions compared to other halo finders, including \Rockstar (see
Paper II). Based on these considerations, we trust the (current
calibration of the) model for the study of MW-size hosts.  If future
simulations confirm steeper power-law slopes at Galactic mass scales,
then it means that some of our model parameters, such as $\zeta$ and
$\bar{A}$, must have some mass dependence
\footnote{For example, we find that with $\zeta=0.19$ and
  $\bar{A}=1.29$ the model accurately reproduces the ELVIS results.},
for which we thus far see no indication in the Bolshoi simulation that
covers 2-3 orders of magnitude in host halo mass.
\begin{figure*}
\centerline{\psfig{figure=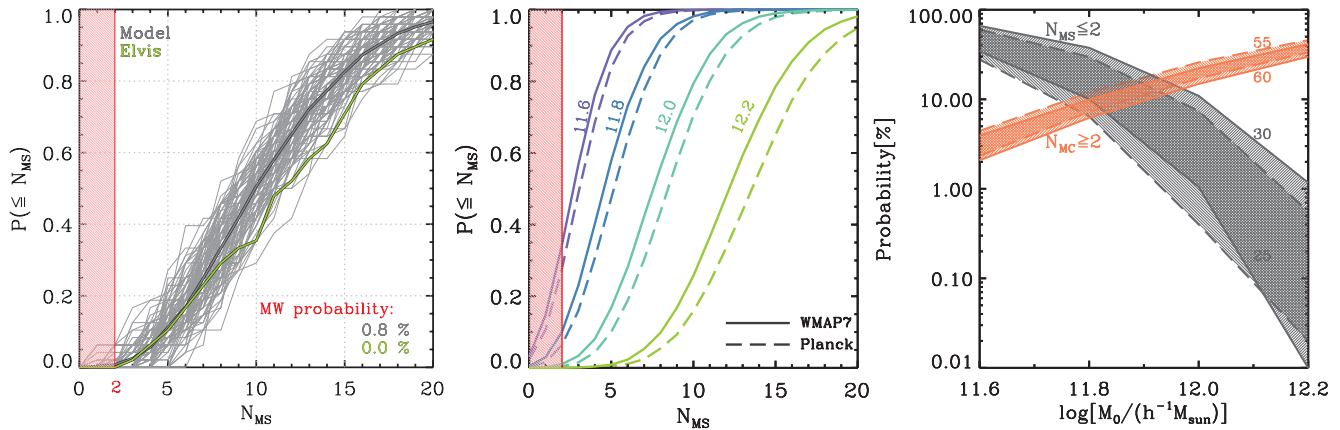,width=1.0\hdsize}}
\caption{ {\it Left-hand panel}: the fraction of host haloes $\leq
  N_{\rm MS}$ massive subhaloes for the 4800 model realizations of
  ELVIS-size haloes (thick, grey line), the actual ELVIS simulation
  suite (green), and the 100 mock ELVIS suites (thin, grey lines).
  Realizations with $N_{\rm MS}\le2$ are considered as MW-consistent.
  {\it Middle panel}: same as the left -hand panel, but for 10,000
  realizations of different halo masses and cosmologies, as indicated.
  {\it Right-hand panel}: the probabilities, i.e., fraction out of
  10,000 realizations, of having no more than two massive subhaloes
  (grey) and no less than two Magellanic-Cloud analogs (orange) as a
  function of halo mass.  The upper and lower bounds of each band
  correspond to different threshold $\Vmax$ values, as indicated,
  while solid and dashed curves indicate results for the WMAP7 and
  Planck cosmologies, respectively.}
\label{Fig:MassiveFailure}
\end{figure*}
%


\section{Assessing the Too Big to Fail Problem} 
\label{Sec:Severity}

In this section, we use the model introduced above to gauge the
severity of TBTF in the massive subhaloes formulation
(\S\ref{Sec:MassiveFailure}), the gap formulation
(\S\ref{Sec:VmaxGap}), and the density formulation
(\S\ref{Sec:Density}).  We use our model to generate 100 realizations
of the ELVIS suite as follows. Adopting the same WMAP7 cosmology as
used for the ELVIS simulations, for each halo mass in the ELVIS suite
of 48, we generate 100 model realizations with a mass resolution of
$m/M_0 = 10^{-5}$. The resulting ensemble of 4800 MW-size host haloes
($M_0=10^{12.08\pm0.23}\Msunh$) is then split in 100 mock ELVIS
suites, each with exactly the same distribution of halo masses. We use
this set below to compare our model predictions with those from the
ELVIS suite, and to gauge the suite-to-suite variation of the various
TBTF statistics. In addition, in order to probe the dependence on halo
mass and cosmological parameters, we also construct ensembles of
10,000 model realizations each for several values of $M_0$, and for a
few different cosmologies.

When comparing model predictions to data, we use the $\Vmax$ values
for the MW and its satellites listed in Table~1. These are compiled
from Xue \etal (2008), van der Marel \&
  Kallivayalil \etal(2014), Kallivayalil \etal (2013), Kuhlen (2010),
and Boylan-Kolchin \etal (2012).  Most of the
  $\Vmax$ values for the dSphs are taken from Kuhlen (2010) and
  Boylan-Kolchin \etal (2012). Both studies used similar methodology
  which we briefly describe in what follows.

The only kinematic information available for the
MW dSphs are the line-of-sight velocities of individual stars, which
constrain the dynamical mass, $M(<r)$, enclosed within some radius,
$r$, that is typically much smaller than $\Rmax$ (e.g., Strigari
\etal 2008; Wolf \etal 2010). In order to infer the corresponding
$\Vmax$, Kuhlen (2010) and Boylan-Kolchin \etal (2012) assign
weights to subhaloes in the Via Lactea II and Aquarius simulations,
respectively, based on how closely they match the measured $M(<r)$.
The corresponding dSph is then assigned the weighted-average $\Vmax$
of these subhaloes.  A crucial underlying assumption of this method
is that the subhaloes in the zoom-in simulations used are
representative of the $\Lambda$CDM subhalo populations of MW sized
haloes. The validity of this assumption is challenged by the
dramatic halo-to-halo variance of subhalo populations.  Fortunately,
for the few dSphs covered by both Kuhlen (2010) and Boylan-Kolchin
\etal (2012), the inferred $\Vmax$ values are mutually consistent
within the errors. When available, we use the more recent results of
Boylan-Kolchin \etal (2012), since they are based on a larger sample
of six host haloes, as compared to only one in the case of Kuhlen
(2010).

For the few dwarfs without published $\Vmax$ constraints (Sgr dwarf, 
Bootes II, Segue II, Bootes I, and Leo V), we use the relation 
$\Vmax = 2.2\sigma_{\rm LOS}$ advocated by Rashkov \etal (2012), 
and the published stellar line-of-sight velocity dispersion 
$\sigma_{\rm LOS}$ (McConnachie 2012 and references therein). 
The resulting $\Vmax$ values are indicated in brackets in Table~1. 
Note that we do not include the Canis Major and Bootes III stellar 
overdensities in this list, as we consider them already disrupted. We 
emphasize, though, that including them in the inventory of MW satellites 
has no impact on any of our results and/or conclusions.

\subsection{The Abundance of Massive Subhaloes} 
\label{Sec:MassiveFailure}

The TBTF problem was originally expressed as a tension between the
rotation curves of the $\sim$10 most massive subhaloes in MW-size host
haloes and the kinematics data of the $\sim$10 brightest MW dwarf
spheroidals (hereafter dSphs).  The MW dSphs all have stellar
kinematics consistent with $\Vmax \lta 25\kms$, while the subhaloes
have $\Vmax \gta 25\kms$.  Therefore, several studies have used the
abundance of subhaloes with $\Vmax$ greater than some threshold value
(typically in the range of 25-30$\kms$) as a measure of the TBTF
severity (e.g., Boylan-Kolchin \etal 2011, 2012; Wang \etal 2012;
Garrison-Kimmel \etal 2014b).  In particular, Garrison-Kimmel
\etal(2014b) define `massive failures' as subhaloes that started out
massive ($\Vacc > 30\kms$) and remain massive ($\Vmax > 25\kms$) to
the present day. The argument is that such subhaloes have potential
wells in which galaxy formation is expected to `succeed', to the
extent that the absence of a satellite galaxy signals a `massive
failure'. We refrain from this nomenclature, as it leads to confusion
when addressing the LMC and SMC; instead, we simply refer to subhaloes
with $\Vacc > 30\kms$ and $\Vmax > 25\kms$ as `massive subhaloes'.
\begin{table}\label{Tab:MWsatellites}
\caption{Maximum circular velocity of MW and its satellites.}
\begin{center}
\begin{tabular}{lcc}
\hline\hline
Object            & $\Vmax$        & Reference \\
                  &   [$\kms$]          &           \\
\hline
Milky Way         & $170.0 \pm 15.0$    & [1] \\
LMC                &   $91.7 \pm  18.8$    & [2] \\
SMC               &  $60.0 \pm  5.0$    &  [3]\\
Sagittarius       & ($25.1 \pm  1.5$)   &     \\
Bootes II         & ($23.1 \pm 16.3$)   &     \\
Draco             &  $20.5^{+4.8}_{-3.9}$ & [5]\\
Ursa Minor        &  $20.0^{+2.4}_{-2.2}$ &  [5]\\
Fornax            &  $17.8\pm0.7$       &  [5]\\
Sculptor          &  $17.3^{+2.2}_{-2.0}$ &  [5]\\
Leo I             &  $16.4^{+2.3}_{-2.0}$ & [5] \\
Ursa Major I      &  $14^{+3}_{-1}$      & [4]\\
Ursa Major II     &  $13^{+4}_{-2}$      & [4]\\
Leo II            &  $12.8^{+2.2}_{-1.9}$ & [5] \\
Sextans           &  $11.8^{+1.0}_{-0.9}$ & [5]\\
Canes Venatici I  &  $11.8^{+1.3}_{-1.2}$ & [5]\\
Carina            &  $11.4^{+1.1}_{-1.0}$ & [5]\\
Canes Venatici II &  $11^{+2}_{-2.1}$     & [4]\\
Hercules          &  $11^{+3}_{-1.6}$     & [4]\\
Segue I	          &  $10^{+7}_{-1.6}$     & [4]\\
Coma Berenices    &  $9.1^{+2.9}_{-0.9}$  & [4]\\
Willman 1         &  $8.3^{+2.7}_{-0.8}$  & [4]\\
Leo V             & ($8.1^{+5.1}_{-3.1}$) &     \\
Segue II          & ($7.5^{+5.5}_{-2.6}$) &     \\
Bootes I          & ($5.3^{+2.0}_{-1.1}$) &     \\
Leo IV            &  $5.0^{+2.2}_{-0.8}$  & [4]\\
\hline\hline
\end{tabular}
\end{center}
\medskip
\begin{minipage}{\hssize}
The references for the values of $\Vmax$ listed in Column (3) are: 
[1] Xue \etal (2008); 
[2] van der Marel \& Kallivayalil (2014);
[3] Kallivayalil \etal (2013); 
[4] Kuhlen (2010); and 
[5] Boylan-Kolchin \etal (2012). 
Values in brackets are inferred from the empirical relation for MW dSphs, 
$V_{\rm max}=2.2\sigma_{\rm LOS}$ (Rashkov \etal 2012), with 
$\sigma_{\rm LOS}$ measurements from McConnachie (2012) and 
references therein.
\end{minipage}
\end{table}

The left-hand panel of Fig.~\ref{Fig:MassiveFailure} plots the
cumulative distribution of the number of massive subhaloes, $N_{\rm
  MS}$, per host halo, in each of the 100 mock Elvis suites (thin,
gray lines). For comparison, the thick, green line shows the results
from the actual ELVIS suite. Typically, the $N_{\rm MS}$ distribution
is broad, ranging from $N_{\rm MS}=0$ to $\sim 30$. The {\it median}
ranges from 8 to 13, which nicely brackets the median of the ELVIS
suite, which is 11.  The red, vertical line corresponds to $N_{\rm MS}
= 2$ and indicates the number of massive subhaloes around the Milky
Way, which correspond to the LMC and SMC.  Based on the 4800 model
realizations (which sample exactly the same host halo masses as the
ELVIS suite), only $0.8\%$ of the MW-sized haloes have no more than
two massive subhaloes. The suite-to-suite variance of that percentage
ranges from 0\% to 6\%, indicating that a set of 48 host haloes is
insufficient for a meaningful evaluation of the TBTF problem.

\subsubsection{Mass and Cosmology Dependence}
\label{Sec:failuresMassdep}

The middle panel of Fig.~\ref{Fig:MassiveFailure} plots the cumulative
distributions of $N_{\rm MS}$ for four different host halo masses
$\log[M_0/(h^{-1}\Msun)] = 11.6$, $11.8$, $12.0$ and $12.2$, and for
two different cosmologies; `WMAP7' and `Planck'. The latter has
$(\Omega_{\rmm,0}, \Omega_{\Lambda,0}, \Omega_{\rmb,0}, h, \sigma_8,
n_\rms) = (0.3175, 0.6711, 0.0486, 0.6825, 0.8344, 0.9624)$, which are
the values inferred from the Planck cosmic microwave background data
(Planck Collaboration \etal 2014). Each is computed using a sample of
10,000 model realizations. Note how decreasing the mass of the host
halo results in a larger fraction of realizations that is consistent
with the Milky Way in that $N_{\rm MS} \leq 2$.  Changing the
cosmology from `WMAP7' to `Planck' causes a slight reduction in the
MW-consistent-fraction. We stress, though, that these probabilities
are very sensitive to the exact definition of `massive
subhalo'. Following Garrison-Kimmel \etal (2014b), in the left and
middle panels of Fig.~\ref{Fig:MassiveFailure} we have defined massive
subhaloes as having $\Vacc \geq 30\kms$ and $\Vmax \geq 25\kms$. In
order to gauge the sensitivity to these exact definitions, we have
repeated the inventory of massive subhaloes changing the requirement
for the present-day maximum circular velocity to $\Vmax \geq 30\kms$.
This drastically increases the MW-consistent-fraction, as is evident
from the right-hand panel of Fig.~\ref{Fig:MassiveFailure}, which
summarizes our results. The gray band indicates the probability
$P(N_{\rm MS} \leq 2)$ as function of halo mass. Solid and dashed
lines correspond to the WMAP7 and Planck cosmologies, whereas the
upper and lower bounds correspond to defining massive subhaloes as
obeying $\Vmax \geq 30\kms$ or $25\,\kms$, respectively. As is evident,
one can boost the MW-consistent-fraction to well over 10\% by
simply reducing the mass of the MW halo to $<
10^{11.8}\Msunh$.

These results are in qualitative agreement with Wang \etal (2012),
who, based on an investigation of the Millenium-II simulation, argued
that the fraction of host haloes having three or fewer subhaloes with
$\Vmax > 30\kms$ increases from $\la 5\%$ to $\sim 40\%$ as $M_0$
decreases from $10^{12.15}\Msunh$ to $10^{11.85}\Msunh$. Similar
results were also reported by Vera-Ciro \etal (2013), who, using a
semi-analytical model of galaxy formation find that if the Milky-Way
haloes are scaled down to $M_0 = 10^{11.75}\Msunh$, the number of
satellites brighter than Fornax can be lowered from order of 10 to
2-5. 
\begin{figure}
\centerline{\psfig{figure=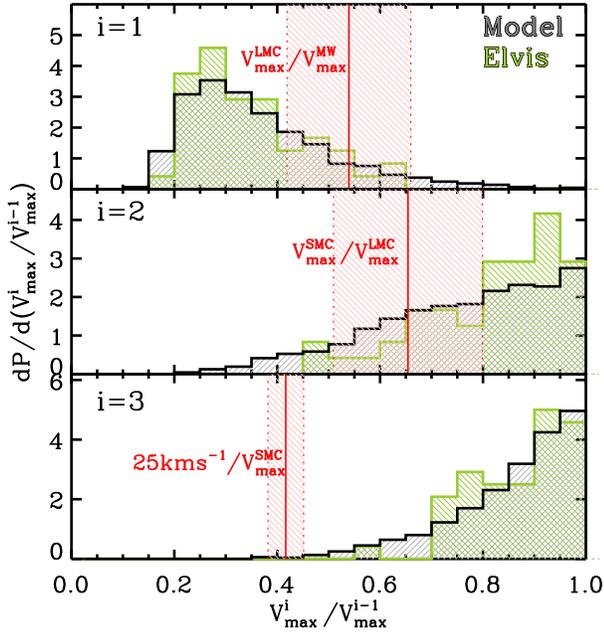,width=0.5\hdsize}}
\caption{The distributions of the $\Vmax$-ratios between the $i^{\rm
    th}$ and $(i-1)^{\rm th}$ subhalo (in descending order of $\Vmax$,
  where $i=0$ refers to the host halo itself), for $i=1,2,3$ (from top
  to bottom).  The grey distributions are obtained from the 4800 model
  realizations of the ELVIS-size haloes, and the green histograms are
  the ELVIS results.  The vertical, red lines mark the corresponding
  MW values, with the red bands indicating the uncertainties due to
  the errors on the $\Vmax$ measurements of the MW, LMC and SMC (see
  Table~1).}
\label{Fig:VmaxRatio}
\end{figure}

To summarize, according to the massive subhalo formulation the TBTF
problem can be significantly alleviated by simply lowering the mass of
the Milky Way halo to slightly below $10^{12} \Msunh$, which is well
within current bounds (e.g., Xue \etal 2008; Kafle \etal
2014). However, this alleged solution ignores an important
observational fact, namely that the two massive subhaloes of the Milky
Way (i.e., the LMC and SMC) actually have fairly large (inferred)
values for $\Vmax$. According to van der Marel \& 
Kallivayalil \etal(2014) and Kallivayalil \etal (2013), the maximum
circular velocity of the LCM and SMC are  
$91.7 \pm 18.8\kms$ and $60.0\pm 5.0\kms$, 
respectively. Hence, the MW seems to have two subhaloes
with $\Vmax \gta 60 \kms$. The orange band in the right-hand panel of
Fig.~\ref{Fig:MassiveFailure} shows the probability that the number,
$N_{\rm MC}$, of Magellanic-Cloud-like systems is larger than or equal
to two. The latter are defined as subhaloes with $\Vmax \geq 55\kms$
(upper solid and dashed curves) or $\geq 60\kms$ (lower solid and
dashed curves). Clearly, the probability $P(N_{\rm MC} \geq 2)$
decreases with decreasing halo mass. Hence while lowering the mass of
the MW halo reduces its abundance of massive subhaloes with $\Vmax
\gta 25\kms$, it also makes it less likely that they have $\Vmax$
values in agreement with the Magellanic clouds.

\subsection{Gap Statistics} 
\label{Sec:VmaxGap}

Based on the above findings, the real tension between the Milky Way
and a simulated MW-size halo is that the third most massive satellite
has a surprisingly small $\Vmax$ of $\sim 25\kms$ (irrespective of
whether this third system is Draco, Ursa Minor, Fornax or the Sgr dwarf)
compared to that of its second most massive satellite, the SMC, which
seems to have $\Vmax \sim 60\kms$. Therefore, a more specific
formulation of the TBTF problem is the existence of a $\Vmax$ gap
between the second and the third most massive MW satellites.

Fig.~\ref{Fig:VmaxRatio} plots the distributions of the $\Vmax$
ratio between the $i^{\rm th}$ and $(i-1)^{\rm th}$ subhaloes,
rank-ordered by $\Vmax$, for $i=1,2$, and 3. Here the $0^{\rm th}$
subhalo corresponds to the host halo itself.  The model predictions
for the 100 mock ELVIS suites and the actual ELVIS simulation results
are in excellent agreement for all three cases.  We take the model
distributions as benchmarks to gauge how (a)typical the MW halo is.
Using the $\Vmax$ values listed in Table~1, the ratio of the first
subhalo to the host halo, $\Vmax^{1{\rm st}} / \Vmax^{0{\rm th}}$, for
the MW is $0.54\pm 0.12$ 
(indicated as the vertical, hatched band in
the upper panel of Fig.~\ref{Fig:VmaxRatio}), well within the
theoretical range.  Similarly, the $\Vmax$-ratio of the second to
first ranked subhalo (middle panel) for the MW is $0.65\pm 0.14$, again
in good agreement with the theoretical predictions. Hence, it is
clear that the subhaloes that host the two Magellanic clouds are
perfectly common. The TBTF problem is revealed in the third panel:
assuming $\Vmax = 25\kms$ for the third MW satellite, and the
aforementioned $\Vmax$ measurement for the SMC, the $\Vmax$-ratio for
$i=3$ for the MW is $0.42\pm0.04$, which is small compared to the
model prediction whose 95 percent confidence interval ranges
from 0.56 to 0.99.
\begin{figure*}
\centerline{\psfig{figure=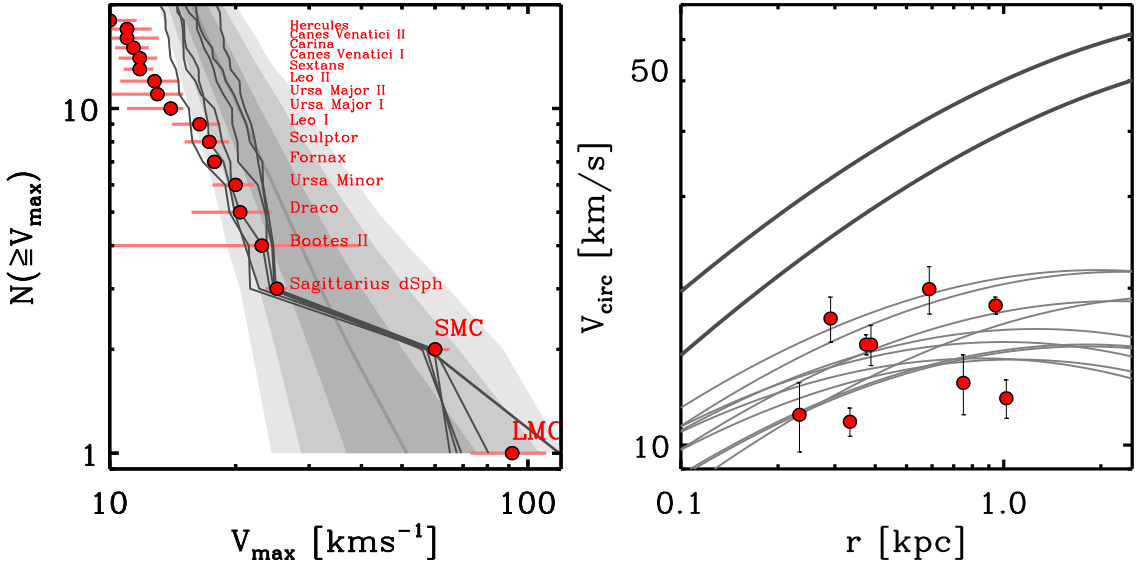,width=\hdsize}}
\caption{Rare model realizations with a $\Vmax$ gap consistent with
  that observed in the MW. {\it Left-hand panel:} grey, solid lines
  show the the cumulative $\Vmax$ functions for the only six
  realizations (out of 10,000 haloes with $M_0=10^{11.8}\Msunh$) that
  reveal a MW-consistent $\Vmax$ gap. For comparison, the solid
  circles with error bars show the actual MW data (see Table~1).  The
  median, and the 68, 95, and 99.7 percentiles of the 10,000 model
  realizations are indicated in a shaded background with progressively
  lighter grey.  {\it Right-hand panel:} the rotation curves of the
  first 12 subhaloes with the highest $\Vmax$ values in one of those
  six realizations (the other five cases look very similar). The two
  rotation curves with a thicker line-style indicate the
  Magellanic-Cloud analogs, while symbols with error bars represent
  the nine brightest MW dSphs (data taken from Wolf \etal 2010).  }
\label{Fig:CircularVelocityProfiles}
\end{figure*}

Interestingly, the $\Vmax^{3{\rm rd}}/\Vmax^{2{\rm nd}}$-distribution
predicted by the model extends down to values well below that of the
MW, indicating that it is possible, albeit rare, for dark matter
haloes to reveal a $\Vmax$-gap comparable to that in the MW. To make
this more quantitative, we construct an ensemble of 10,000 model
realizations for $M_0 = 10^{11.8} \Msunh$ in the WMAP7 cosmology and
identify realizations with $\Vmax^{3{\rm rd}} \leq 25\kms$ and
$\Vmax^{2{\rm nd}} \geq 55\kms$. Among the 10,000 realizations, we
only find 6 ($0.06\%$) that meet these criteria\footnote{We also
  repeated this exercise for host haloes with $M_0 = 10^{12.0}
  \Msunh$, which resulted in only a single candidate.}. The left-hand
panel of Fig.~\ref{Fig:CircularVelocityProfiles} compares the
cumulative subhalo velocity functions for these six MW-consistent
cases (solid lines) with that of the actual MW satellites (red symbols
with error-bars; data taken from Table~1). Note that three of the six
model realizations mimic the MW extremely well down to 
$\Vmax \sim 15\kms$. For
  smaller $\Vmax$ all models predict far more satellites than
  observed. We emphasize though, that the inventory of MW satellites
  is incomplete at the low $\Vmax$ end.  Tollerud \etal (2008) have
  demonstrated that the luminosity function of MW satellites suffers
  from incompleteness for $M_V \ga -9$\footnote{The eight new MW
    dwarfs reently disovered in the Dark Energy Survey (DES
    Collaboration \etal 2015) are indeed all much fainter than this).}.
  Since there are ten MW satellites brighter than this, we estimate
  that the MW inventory is roughly complete down to $\Vmax \sim
  15\kms$.  Although we acknowledge that the rank-order in $M_V$ is
  not equal to that in $\Vmax$, we argue that the discrepancy between
  model and data for $\Vmax < 15\kms$ is most likely a manifestation
  of incompleteness in the data.

The right-hand panels of Fig.~\ref{Fig:CircularVelocityProfiles} plot
the circular velocity curves for the 12 subhaloes with the largest
$\Vmax$ in one of our model realizations, randomly chosen from the set
of six shown in the left-hand panel (results for the other 5 are
similar).  Following Garrison-Kimmel \etal (2014b), these circular
velocity profiles are computed using the $\Vmax$ and $\Rmax$ values
for each subhalo as predicted by the model (see
\S\ref{Sec:ModelOverview}), and assuming that subhaloes follow an
Einasto profile (Einasto 1965)
\begin{equation} \label{Einasto}
\rho(r) = \rho_{-2} \, \exp\left[{-2 \over \alpha}\left\{
\left({r \over r_{-2}}\right)^{\alpha} - 1\right\} \right]
\end{equation}
where $r_{-2}$ is the radius at which the logarithmic slope of the
density distribution is equal to $-2$, $\rho_{-2} = \rho(r_{-2})$ and
$\alpha$ is a shape parameter. In computing our model circular
velocity curves we have adopted $\alpha = 0.18$, which is the typical
value for subhaloes in the Aquarius simulation (Springel \etal 2008)
\footnote{Subhaloes in the Aquarius simulation cover the range $0.15 <
  \alpha < 0.30$. Note of our results change significantly if we vary
  $\alpha$ over this range.}.  For comparison, the red circles with
error-bars indicate the circular velocity measurements at the
half-light radius, $V_{\rm circ}(r_{1/2})$, of the 9 brightest MW
dSphs (Wolf \etal 2010). Note how this model predicts two Magellanic
cloud analogs, with $\Vmax = 68\kms$ and $57\kms$ respectively, whose
circular velocity curves clearly stand out, with a pronounced gap to
the subsequent subhaloes. Note also how the circular velocity curves
of these subsequent subhaloes are statistically consistent with the
data for the MW dSphs.

Hence, based on the gap statistics, we conclude that, within the
$\Lambda$CDM paradigm, one can find MW-sized haloes with subhalo
statistics in excellent agreement with our best current understanding
of the Milky Way. However, such haloes seem to be extremely rare.  In
order to quantify this in more detail, we now investigate the gap
statistic in more detail, and as function of host halo mass.

\subsubsection{Dependence on Host Halo Mass}
\label{Sec:gapMassdep}

\begin{figure*}
\centerline{\psfig{figure=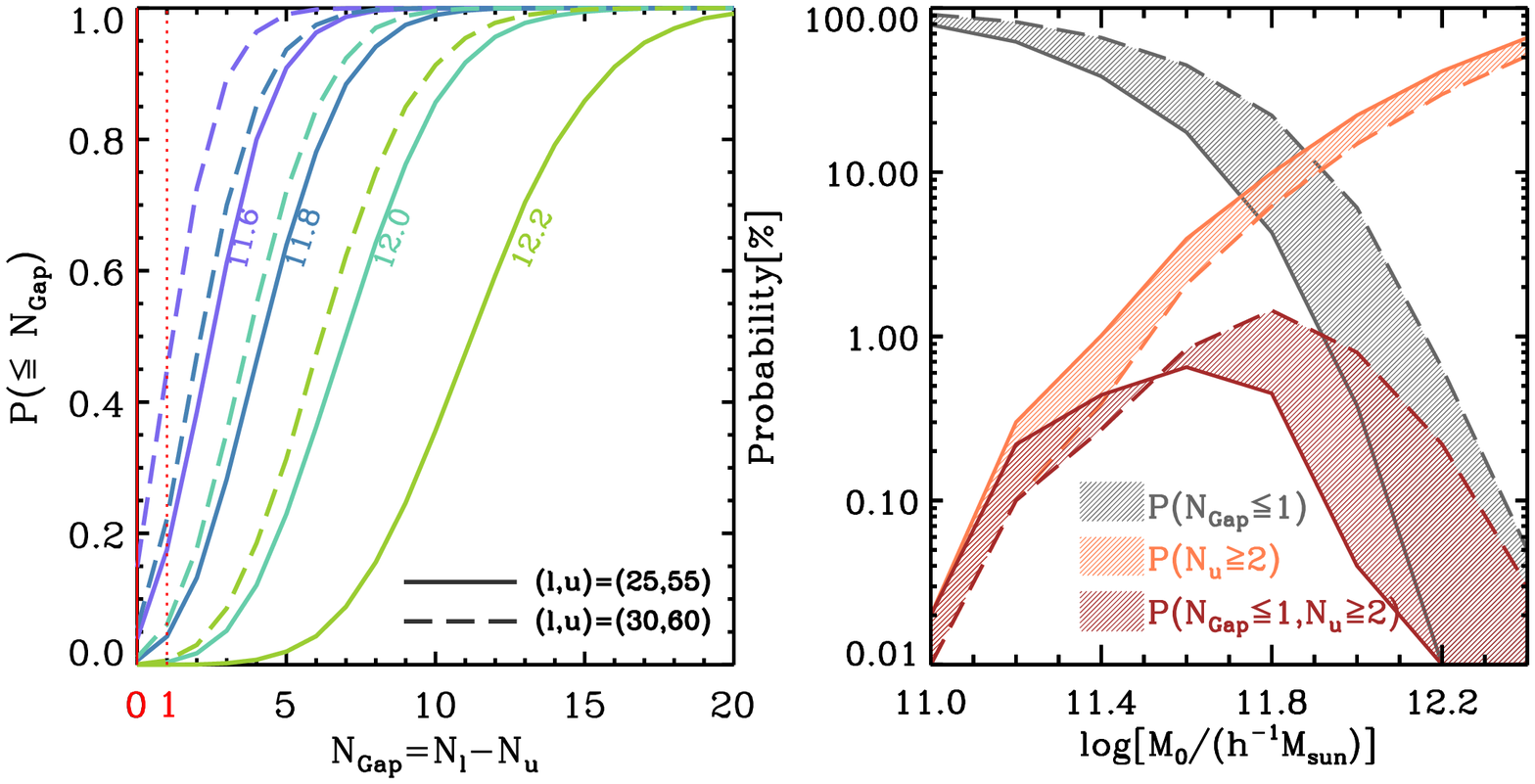,width=0.8\hdsize}}
\caption{ {\it Left-hand panel}: the fraction of model realizations
  with no more than $N_{\rm Gap}$ subhaloes with $\Vmax\in(25,55)\kms$
  (solid lines) or $\Vmax\in(30,60)\kms$ (dashed lines), for different
  host halo masses as indicated (value reflects
  $\log[M_0/(h^{-1}M_{\odot})]$).  {\it Right-hand panel}: the
  probabilities of having no more than one subhalo in the $\Vmax$ gap,
  $P(N_{\rm Gap}\le1)$, no less than two Magellanic-Cloud analogs,
  $P(N_\rmu\ge2)$, and both $N_{\rm Gap}\le1$ and $N_\rmu\ge2$,
  $P(N_{\rm Gap}\le1,N_\rmu\ge2)$, as a function of host halo
  mass. The solid and dashed curves correspond to different threshold
  $\Vmax$ values, as indicated in the left-hand panel.  }
\label{Fig:GapStat}
\end{figure*}

The left-hand panel of Fig.~\ref{Fig:GapStat} plots the cumulative
distributions of $N_{\rm gap} \equiv N_\rml - N_\rmu$. Here $N_\rml$
and $N_\rmu$ are the numbers of subhaloes with $\Vmax$ larger than
some {\it lower} and {\it upper} limit, respectively. For the solid
lines we set these lower and upper limits to be $25 \kms$ and $55
\kms$, respectively. Thus defined, $N_{\rm gap}$ is the number of
subhaloes in the range $25 \kms < \Vmax < 55 \kms$. Different colors
correspond to different host halo masses, as indicated.  For each host
halo, we compute $P(\leq N_{\rm gap})$ using 10,000 model realizations
for a WMAP7 cosmology. Taking the best-fit values for $\Vmax$ of each
MW satellite galaxy, as listed in Table~1, and ignoring their
uncertainties, the Milky Way has $N_{\rm gap} = 1$ (the Sgr dSph, for
which $\Vmax = 25.1 \kms$).  As is evident from the left-hand panel of
Fig.~\ref{Fig:GapStat}, the probability that a $\Lambda$CDM halo has
$N_{\rm gap} \leq 1$ is extremely small if $M_0 \gta 10^{12} \Msunh$,
but rapidly increases with decreasing $M_0$. This is quantified more
clearly by the solid, black line in the right-hand panel of
Fig.~\ref{Fig:GapStat} which plots $P(N_{\rm gap} \leq 1)$ as function
of host halo mass. We emphasize, though, that these results are very
sensitive to how exactly one defines the gap. Setting the lower and
upper limits of the gap to $30 \kms$ and $60 \kms$, for example,
results in the dashed curves, for which $P(N_{\rm gap} \leq 1)$ is
significantly larger.

The orange, hatched area in the right-hand panel indicates the
probability $P(N_\rmu \geq 2)$, where the upper limit corresponds to
$\Vmax = 55\kms$ (solid line) or $60 \kms$ (dashed line),
respectively. Note that the probability to host two subhaloes
consistent with the two Magellanic Clouds drops below one percent for
$M_0 \lta 10^{11.4} \Msunh$. Hence, to be consistent with the MW
requires that $N_{\rm gap} \leq 1$ and $N_\rmu \geq 2$. The
probability that both constraints are satisfied is indicated by the
red band in the right-hand panel, which plots $P(N_{\rm gap}\leq 1,
N_\rmu \geq 2)$ as function of halo mass.

The dashes lines correspond to a different choice for the lower and
upper limits of $30 \kms$ and $60 \kms$, respectively. Note how the
dashed curves are shifted to the left compared to the solid curves,
indicating that the typical $N_{\rm gap}$ is significantly smaller for
this definition of the $\Vmax$-gap.  Although not shown here, we have
verified that changing cosmology from WMAP7 to Planck has weak
influence on $P(N_{\rm gap} \leq 1)$ and $P(N_\rmu \geq 2)$, similar
to that shown in Fig.~\ref{Fig:MassiveFailure}, and negligible effect
on $P(N_{\rm gap}\leq 1, N_\rmu \geq 2)$.

The gap statistic reveals a stronger tension than the massive-subhalo
count. In particular, at best $\sim 1\%$ of $\Lambda$CDM haloes have a
$\Vmax$-gap comparable to that of the MW, and that is for a MW host
halo mass of $M_0 \simeq 10^{11.8}\Msunh$. This probability drops to
below $0.1\%$ for $M_0 \lta 10^{11.2} \Msunh$ and $M_0 \gta
10^{12.2}\Msunh$. These results are in excellent agreement with the
empirical extrapolations of $N$-body simulation results by Cautun
\etal (2014b).

\subsection{Density of Massive Subhaloes} 
\label{Sec:Density}

The third formulation of TBTF, the one used in the paper by
Boylan-Kolchin \etal (2011) that introduced the TBTF problem, is that
the most massive subhaloes in simulations are too dense to be
consistent with constraints on the MW dSphs.  Here the internal
densities of satellites are usually characterized in terms of their
$\Rmax$ and $\Vmax$ values.
\begin{figure*}
\centerline{\psfig{figure=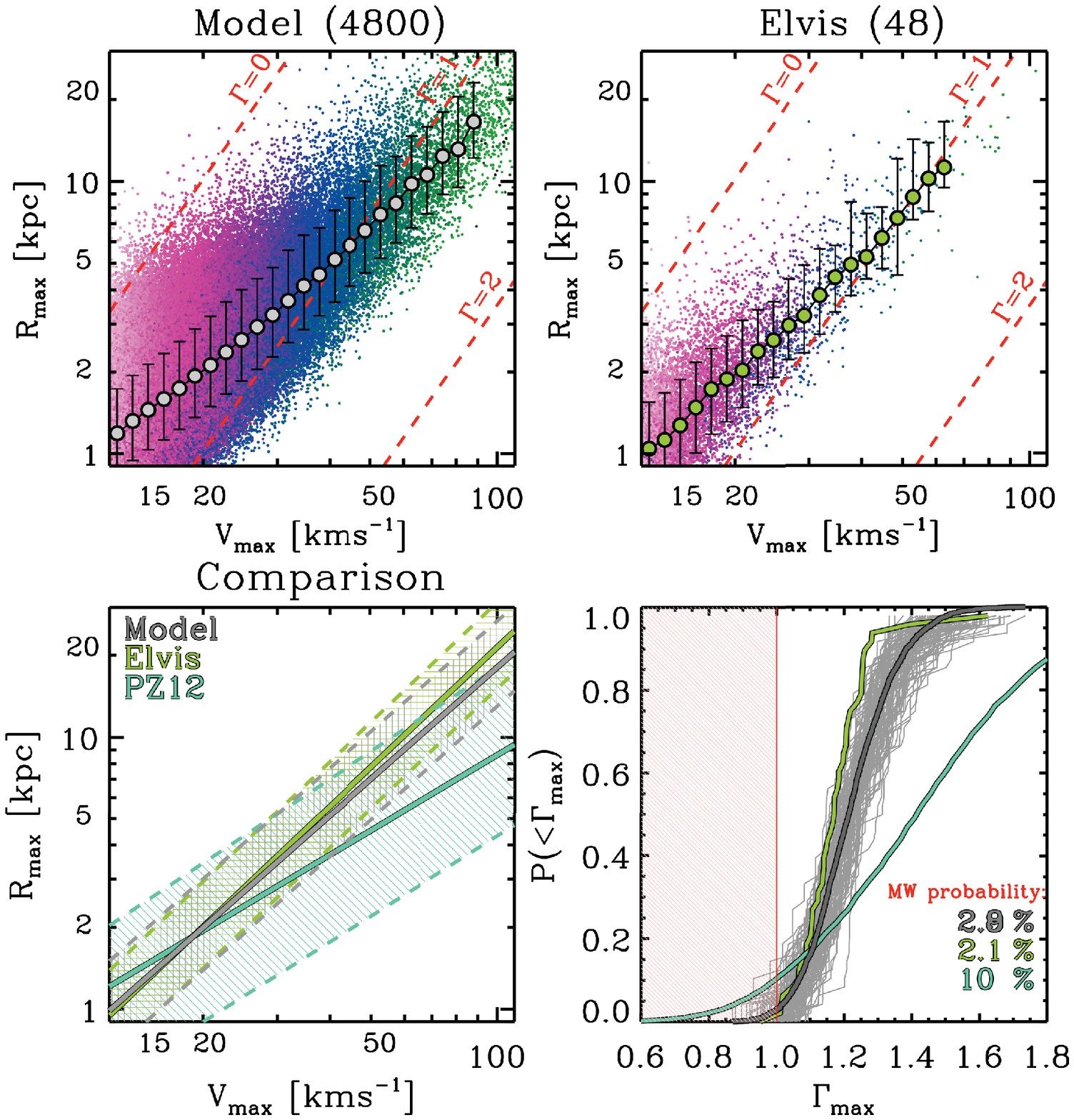,width=\hdsize}}
\caption{Distribution of subhaloes in $\Rmax$-$\Vmax$ space, for the
  4800 model realizations of ELVIS-size haloes ({\it upper, left-hand
    panel}) and the 48 ELVIS haloes ({\it upper, right-hand panel}).
  Individual dots are subhaloes, color coded by $V_{\rm acc}$.
  Circles indicate the median $\Rmax$ at given $\Vmax$ bins, with
  error bars indicating the 16 and 84 percentiles.  The red dashed
  lines are the loci of $\Gamma=0,1$, and 2 [Eq.(\ref{Eq:Gamma})],
  with $0<\Gamma<1$ encompassing the region occupied by MW dSphs.
  {\it Lower, left-hand panel}: comparison of the best-fit
  $\Rmax$-$\Vmax$ relations of our model (grey), the ELVIS simulation
  (green), and the PZ12 model (cyan), with solid and dashed lines
  indicating the median relations and the 16 and 84 percentiles
  respectively. {\it Lower, right-hand panel}: corresponding cumulative
  $\Gamma_{\rm max}$ distributions. The red-shaded region corresponds
  to $\Gamma_{\rm max}<1$ and is considered `MW consistent'.  }
\label{Fig:RmaxVmaxDiagram}
\end{figure*}

Measurements of the stellar kinematics of satellite galaxies can put
accurate constraint on their enclosed mass within their half-light
radius (Wolf \etal 2010). These in turn, constrain a degenerate
combination of $\Rmax$ and $\Vmax$ (e.g., Zentner \&
Bullock 2003; Boylan-Kolchin \etal 2011). This motivated PZ12 to
define a density proxy, $\Gamma$, as a linear combination of
$\log(\Rmax)$ and $\log(\Vmax)$;
\begin{equation} \label{Eq:Gamma}
\Gamma \equiv 1+\log(0.0014 \Vmax^{2.2} / \Rmax),
\end{equation}
which increases in a direction approximately orthogonal to the envelop
of the constraint on MW dSphs in the $\log(\Rmax)$-$\log(V_{\rm
  max})$ plane. Thus defined, $\Gamma=1$ corresponds to the 2$\sigma$
upper bound of the region in the $\Rmax$-$\Vmax$ space that is
occupied with MW dSphs. A host halo is said to be
MW-consistent if {\it all} its subhaloes obey $\Gamma \leq 1$, while
the presence of one or more subhaloes with $\Gamma > 1$ constitutes a
manifestation of TBTF.

The upper left-hand panel of Fig.~\ref{Fig:RmaxVmaxDiagram} plots the
subhaloes in the $\Rmax$-$\Vmax$ space for the 4800 model realizations
of our 100 mock ELVIS suites, color-coded according to their value for
$\Vacc$.  The upper right-hand panel shows the same, but this time for
the actual subhaloes in the ELVIS simulation suite of 48 MW-size host
haloes. Filled circles indicate the median $\Rmax$ as a function of
$\Vmax$, with error-bars indicating the 16 and 84 percentiles. Red,
dashed lines indicate the loci of $\Gamma = 0$, $1$ and $2$. The model
predictions are in good agreement with the ELVIS simulation results,
in that both predict a median $\Rmax - \Vmax$ relation corresponding
to $\Gamma > 1$ for $\Vmax \gta 45\kms$.  Note that all observed MW
dwarf spheroidals fall in the range $0 < \Gamma < 1$ (see PZ12).

Following Garrison-Kimmel \etal (2014b), we fit the median $R_{\rm
  max}$-$\Vmax$ relation with a simple power law
\begin{equation} \label{Eq:RmaxVmaxRelation}
\frac{\Rmax}{1{\rm kpc}} = A\left(\frac{\Vmax}{10 {\rm kms}^{-1}}\right)^{p}.
\end{equation}
For the model predictions we find $(A,p)=(0.77,1.37)$, in good
agreement with the ELVIS results, for which $(A,p)=(0.73,1.47)$.  A
comparison of the best-fit $\Rmax$-$\Vmax$ relations of our model, the
ELVIS simulation, and the PZ12 model is shown in the lower left-hand
panel of Fig.~\ref{Fig:RmaxVmaxDiagram}.  The PZ12 model
\footnote{The PZ12 predictions are read off from their published
  figures, which are based on 10,000 realizations of $10^{12}\Msunh$
  haloes in a WMAP7 cosmology. }  predicts significantly larger
scatter and a much shallower slope ($p=0.92$) compared to both our
model and the ELVIS simulation.

The lower right-hand panel of Fig.~\ref{Fig:RmaxVmaxDiagram} plots the
cumulative distributions of $\Gamma_{\rm max} = \max\{\Gamma_i\}$,
where the maximum is taken over all subhaloes in a single host halo.
In the ELVIS suite, only one out of the 48 host haloes (corresponding
to 2.1\%) has $\Gamma_{\rm max} \leq 1$, and is therefore
MW-consistent. This is in good agreement with our model predictions,
for which we find (using a sample of 4800 host haloes) a MW-consistent
fraction of $2.8\%$.  As in \S\ref{Sec:MassiveFailure}, we can use the
100 mock ELVIS suites to infer the suite-to-suite variance.  The
median $\Gamma_{\rm max}$ ranges from 1.1 to 1.3, with a MW-consistent
fraction that varies from 0\% to 15\%.  This is another demonstration
that the sample size of the ELVIS suite is insufficient for an
accurate assessment of TBTF.

As is evident from the lower right-hand panel of
Fig.~\ref{Fig:RmaxVmaxDiagram}, the PZ12 model predicts a $\Gamma_{\rm
  max}$-distribution that is much broader than what is found in the
ELVIS simulation suite or predicted by our model. The relatively large
difference between PZ12 and our model is most likely due to two
reasons.  First, PZ12 construct their merger trees using the
Somerville \& Kolatt (1999) algorithm, which is known to overpredict
merger rates, and result in a halo-to-halo variance of mass assembly
histories that is too large (e.g., Zhang \etal 2008; Jiang \& van den
Bosch 2014a). Our model, instead, relies on the Parkinson, Cole \&
Helly (2008) algorithm, which yields merger trees that are
statistically indistinguishable from those extracted from numerical
simulations. Second, PZ12 use a different model for evolving $\Vmax$
and $\Rmax$ of subhaloes. In particular, whereas we use
Eq.(\ref{Eq:P10}) to compute $\Vmax$ and $\Rmax$, PZ12 simply assume
that $\Vmax \propto m^{1/3}$ and compute $\Rmax$ assuming that dark
matter subhaloes follow an NFW profile. Based on a number of tests, we
conclude that this PZ12 model for computing the structural parameters
of their subhaloes results in a median $\Rmax$--$\Vmax$ relation that
is too shallow, while their merger trees are responsible for
introducing a scatter in the $\Rmax$--$\Vmax$ relation that is too
large, which in turn results in a MW-consistent fraction that is too
high.
\begin{figure*}
\centerline{\psfig{figure=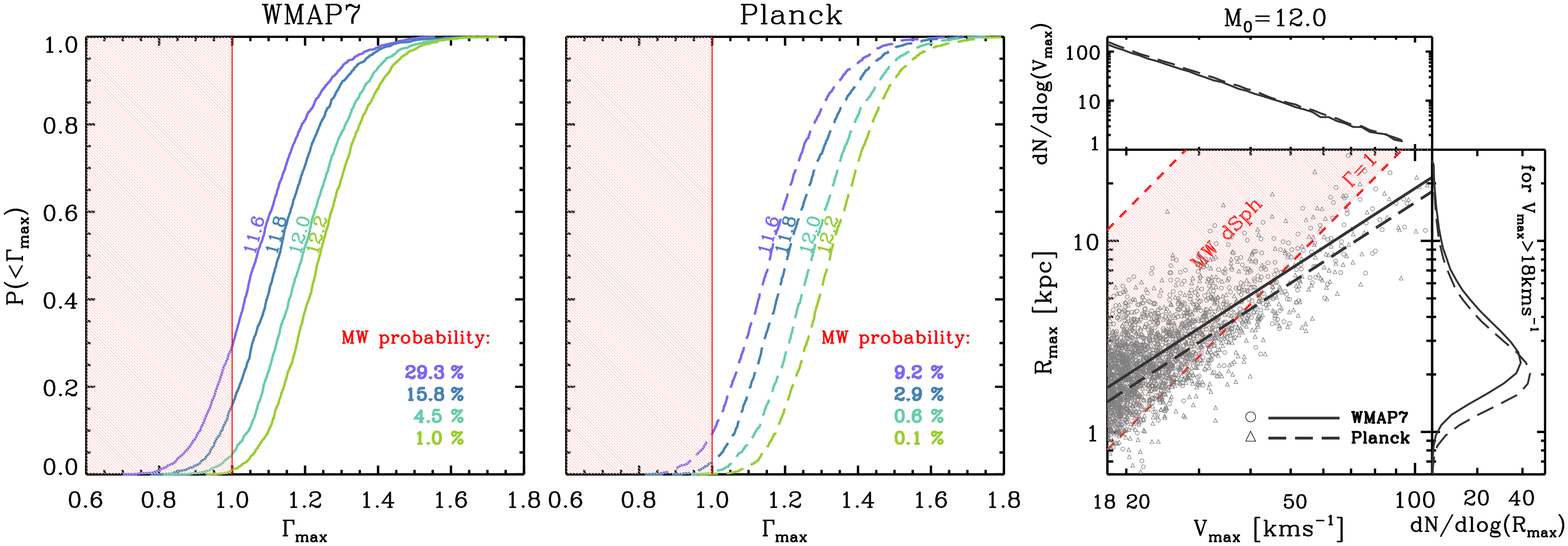,width=1.0\hdsize}}
\caption{Cumulative $\Gamma_{\rm max}$ distributions for different
  halo masses, in the WMAP7 cosmology (left-hand panel), and the
  Planck cosmology (middle panel).  The MW-consistent regime
  ($\Gamma_{\rm max} < 1$) is highlighted in red.  The right-hand
  panel shows the subhalo $\Rmax$--$\Vmax$ relations for a host halo
  mass $M_0 = 10^{12.0}\Msunh$ in the WMAP7 and Planck cosmologies.
  The circles (WMAP7) and triangles (Planck) correspond to a random
  subsample of model realizations, while the solid and dashed lines
  indicate the corresponding median relations. The red shaded band
  indicates the region occupied by MW dSphs.  The top and side panels
  plot the $\Vmax$ and $\Rmax$ distributions for model subhaloes with
  $\Vmax > 18\kms$. Note that subhaloes are expected to be
  significantly denser (i.e., smaller $\Rmax$) in the Planck
  cosmology.}
\label{Fig:Discussion2}
\end{figure*}

We caution that the observational constraint on
  $\Vmax$ and $\Rmax$, expressed as $0 < \Gamma < 1$, is based on the
  assumption that dark matter subhaloes have NFW density profiles.
  Vera-Ciro \etal (2013) have shown that if one instead assumes an
  Einasto profile with $\alpha=0.5$, the constraints on $\Vmax$ and
  $\Rmax$ are significantly altered, to the extent that even the
  densest subhaloes in the Aquarius simulations are now consistent
  with the data (i.e., one would no longer infer a TBTF
  problem). However, subhaloes in numerical simulations typically have
  density profiles that are well fit by an Einasto profile, but with
  $\alpha \sim 0.2$, for which the constraints on $\Vmax$ and $\Rmax$
  are very similar to those obtained assuming an NFW profile. Hence,
  the constraint $0 < \Gamma < 1$ proposed by PZ12 is still valid,
  despite the oversimplified assumption that subhaloes follow an NFW
  profile.

\subsubsection{Mass and Cosmology Dependence}
\label{Sec:densityMassdep}

Fig.~\ref{Fig:Discussion2} investigates the dependence of the
cumulative $\Gamma_{\rm max}$ distribution on halo mass and cosmology.
The left-hand panel plots the results for the WMAP7 cosmology.  The
fraction of realizations with $\Gamma_{\rm max}<1$ increases from 1\%
at $M_0=10^{12.2}\Msunh$ to 29\% at $M_0=10^{11.6}\Msunh$.  Also based
on the WMAP7 cosmology, the PZ12 model predicts the MW-consistent
fractions to be 20\%, 10\% and 10\% for $M_0=10^{11.8}\Msunh$,
$10^{12.0}\Msunh$, and $10^{12.2}\Msunh$ respectively, significantly
higher than our findings of 16\%, 4.5\%, and 0.6\%.

The middle panel of Fig.~\ref{Fig:Discussion2} plots the same results
but now for the Planck cosmology.  The MW-consistent fraction
increases from 0.1\% at $M_0=10^{12.2}\Msunh$ to 9\% at
$M_0=10^{11.6}\Msunh$, significantly smaller that for the WMAP7
cosmology.  Therefore, a relatively small change in cosmological
parameters seems to have a relatively large impact on the
TBTF-statistics in the density formulation.

Polisensky and Ricotti (2014) argued that the cosmology dependence
mainly manifests itself as a change in the $\Rmax$ of subhaloes.  To
test this, the right-hand panel of Fig.~\ref{Fig:Discussion2} plots
the $\Rmax$--$\Vmax$ diagram for the WMAP7 and Planck cosmologies, at
fixed halo mass of $M_0 = 10^{12.0}\Msunh$.  In both cases, the
$\Rmax$--$\Vmax$ relations are well fit by
Eq.~(\ref{Eq:RmaxVmaxRelation}) with $p \simeq 1.4$.  The
normalizations of the best-fit relations, though, are different, with
$A=0.62$ for the Planck cosmology, and $A=0.74$ for the WMAP7
cosmology. This indicates that subhaloes in the Planck cosmology are,
on average, $\sim 20$ percent denser than in the WMAP7 cosmology.  The
top and side panels of the right-hand panel of
Fig.~\ref{Fig:Discussion2} show the average subhalo $\Vmax$ and
$\Rmax$ distributions for all subhaloes with $\Vmax > 18\kms$,
respectively. This clearly shows that the difference in cosmology
predominantly manifests itself as a change in the $\Rmax$
distribution of subhaloes, confirming the results of Polisensky and
Ricotti (2014). 

The cosmology dependence of subhalo densities arises from the
cosmology dependence of the host halo assembly histories: larger
$\Omega_{\rmm,0}$, smaller $\Omega_{\Lambda,0}$ and larger $\sigma_8$,
as in the case of the Planck cosmology compared to the WMAP7
cosmology, all result in earlier (average) formation times for host
haloes of given present-day mass (e.g., van den Bosch 2002;
Giocoli, Tormen \& Sheth 2012). Earlier assembly implies that the host
halo accreted its subhaloes at earlier epochs, when the Universe (and
therefore the dark matter haloes) was denser.


\section{An Alternative Statistic}
 \label{Sec:AltStat}

In the previous section we have used three different statistics that
have been used in the literature to assess the TBTF problem. If we
adopt a MW host halo mass of $M_0 =10^{12} \Msunh$, the inference is
that the MW-consistent fraction ranges anywhere between $\sim 0.1\%$
and $\sim 10\%$, depending on which statistic one uses. Obviously,
this raises the question which is the more meaningful statistic to
use. We believe the answer is basically none of the above, and the
reason is that they either suffer from the "look-elsewhere effect"
(e.g., Gross \& Vitells 2010), and/or disregard certain aspects of the
data.

Both the massive subhaloes formulation and the gap formulation use
statistics that require the identification of one or more particular
values of $\Vmax$; the massive subhaloes formulation considers the
number of subhaloes with $\Vmax$ above some limit, while the gap
statistic is based on the number of satellites between two values of
$\Vmax$. These values are chosen by the user after carefully examining
the data on the MW, in an attempt to find a statistic for which the MW
is least likely. This is a clear example of the look-elsewhere
effect. For example, if the MW satellite system would have revealed a
$\Vmax$-gap between $10\kms$ and $30\kms$, rather than between
$25\kms$ and $55\kms$, this would have raised a similar concern of
being inconsistent with $\Lambda$CDM predictions. Yet, such a gap does
not manifest itself based on the gap statistic used above to assess
TBTF. Hence, rather than asking what the probability is for a gap
between $25\kms$ and $55\kms$, one should ponder about the probability
that a host halo reveals {\it some} gap, not necessarily between these
two exact values. This is also evident from the fact that we have
demonstrated that small changes in the `user-specified' values that
define the gap results in large changes in the MW consistent fraction,
and thus in the inference regarding the severity of TBTF. Ideally,
then, one should use a statistic that is `blind' in that it does not
rely on an examination of the data beforehand.

Another problem with the previous statistics is that both the massive
subhaloes formulation and the density formulation do not properly
account for the Magellanic clouds. We believe this to be a serious
shortcoming, as the Magellanic clouds, by themselves, put a tight
constraint on the mass of the Milky Way host halo (e.g., Busha \etal
2011).

Finally, it is important to realize that no study of TBTF to date has
properly accounted for the observational errors in the $\Vmax$
measurements of the MW satellite galaxies. As we demonstrate below,
this introduces a huge uncertainty on any MW-consistent fraction, and
should be properly taken into account.

Based on these considerations, we devise a new statistic that is
`blind' (i.e., no scale has to be picked upfront), uses all data on
equal footing, and allows for a straightforward treatment of errors in
the $\Vmax$ measurements of individual MW satellites.  Consider two
rank-ordered distributions, ${\calS}_1(x_1, x_2,..., x_N)$ and
${\calS}_2(y_1, y_2,..., y_N)$. In our application, $x_i$ and $y_i$
are the $\Vmax$ values of dark matter subhaloes,
  while ${\calS}_1$ and $\calS_2$ are two different host haloes (i.e.,
  two different model realizations for a host halo of given mass, or a
  model realizations plus the actual $\Vmax$ data for satellite
  galaxies in the MW).  Note that $\calS_1$ and $\calS_2$ have the
same number of elements and that $x_{i+1} \geq x_i$ and $y_{i+1} \geq
y_i$. We now introduce the statistic
\begin{equation}\label{Qstat}
Q \equiv {\sum_{i=1}^N \vert x_i - y_i \vert \over \sum_{i=1}^N (x_i + y_i)}
\end{equation}
which is a measure for the difference between the (cumulative)
distributions of $\calS_1$ and $\calS_2$.  In particular, ${1 \over N}
\sum_{i=1}^N \vert x_i - y_i \vert$ is the absolute value of the area
between the cumulative distributions of $\calS_1$ and $\calS_2$.  We
normalize this area by ${1 \over N} \sum_{i=1}^N (x_i + y_i)$ so that
$Q$ is dimensionless, and insensitive to an overall shift in $x$ and
$y$ (i.e., multiplying $x_i$ and $y_i$ by some factor $f$ leaves $Q$
invariant). Note that $Q = 0$ if $\calS_1 = \calS_2$ (the
distributions are identical), while $Q=1$ if either $\calS_1$ or
$\calS_2$ consists solely of null elements (i.e., $x_i=0$ or $y_i =0$
for all $i=1,2,...,N$).
\begin{figure*}
\centerline{\psfig{figure=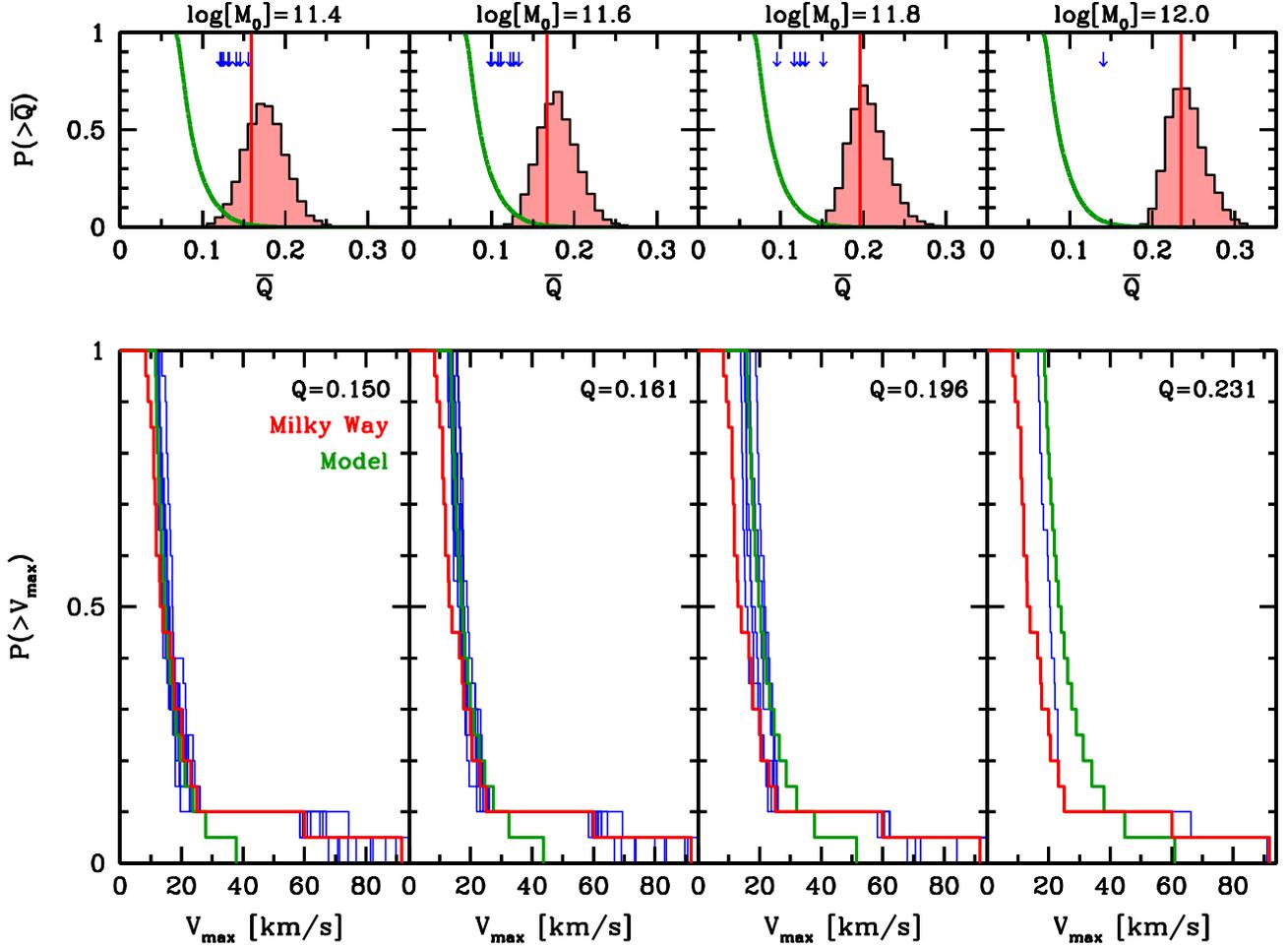,width=0.97\hdsize}}
\caption{{\it Upper panels:} The thick, green curves show the
  cumulative probability distribution, $P(>\bar{Q})$, of the $\bar{Q}$
  statistic defined by Eqs.~(\ref{Qstat})-(\ref{Qbar}), as obtained
  from 10,000 model realizations for dark matter haloes in a WMAP7
  cosmology. Different panels correspond to different host halo
  masses, as indicated. The solid, red line indicates $\bar{Q}_{\rm
    MW}$, while the red-shaded histogram is the distribution of
  $\bar{Q}'_{\rm MW}$ from 10,000 Monte-Carlo realizations of the
  $\Vmax$-distribution of MW satellites obtained by independently
  drawing $\Vmax$ values from their respective error
  distributions. Blue, downward pointing arrows indicate the $\bar{Q}$
  values for (some of) the model realizations that reveal a
  $\Vmax$-gap similar to the MW [i.e., $N(\Vmax>25\kms) =
  N(\Vmax>55\kms) = 2$]. {\it Lower panels:} Cumulative $\Vmax$
  distributions for the MW (solid, red line), the median of the 10,000
  model realizations (sold, green line), and (some of) the model
  realizations with a $\Vmax$-gap similar to the MW. The $Q$ value
  corresponding to the comparison of the red and green curves is
  indicated in the top-right corner of each panel.}
\label{Fig:Qstat}
\end{figure*}
Note that this $Q$-statistic is similar to the Kolmogorov-Smirnov
test, which measures the maximum value of the absolute difference,
$d_{\rm KS}$, between two cumulative distributions.  However, the
KS-test is not well suited to characterize differences in the tails of
two distributions; it mainly is sensitive to finding differences in
the median. We therefore opted to use the statistic $Q$ instead, which
has equal sensitivity throughout the distributions. In adopting $Q$ to
assess TBTF, each individual dark matter host halo has a corresponding
distribution $\calS$, in which the elements are the rank-ordered
values of $\Vmax$ for the $N$ subhaloes with the largest $\Vmax$
values.

In order to turn the $Q$ statistic into a probability measure, we
proceed as follows. Given $K = 10,000$ model realizations, for a given
host halo mass, $M_0$, and a given cosmology, we first compute the
values $Q_{ij}$ for each pair $\{\calS_i,\calS_j\}$ (with $i,j =
1,2,...,K$), where $\calS_i$ is the rank-ordered distribution of the
$N$ largest $\Vmax$ values for model realization $i$. Next we compute
the average
\begin{equation}\label{Qbar}
\bar{Q}_i = {1 \over K-1} \sum_{j\ne i} Q_{ij}
\end{equation}
for each of the 10,000 realizations. Finally, we compute the $K$
values of $Q_{{\rm MW},i}$ by comparing the $\Vmax$ distribution of
the MW to that of each of the 10,000 model realizations, which yields
\begin{equation}
\bar{Q}_{\rm MW} = {1 \over K} \sum_{i=1}^K Q_{{\rm MW},i}
\end{equation}
Using the distribution of 10,000 $\bar{Q}_i$ values, we can now
compute $P(>\bar{Q}_{\rm MW})$, the probability that the $\Lambda$CDM
cosmology yields host haloes with $\bar{Q}$ values as large as that of
the MW. In what follows we refer to this probability as $\calP_{\rm
  MW}$.

Finally, we can easily adopt the $Q$ statistic to also account for
observational errors in the $\Vmax$ values of the MW satellites.
First we construct $N_{\rm MC} = 10,000$ Monte-Carlo realization of
the $\Vmax$-distribution of MW satellites by independently drawing
$\Vmax$ values from their respective error distributions. For
simplicity, we assume that the error-distributions for $\Vmax$ are
Gaussian, with mean and standard deviation equal to the values listed
in Table~1. If the $\Vmax$ values quoted in Table~1 has separate upper
and lower bounds, we assume that the standard deviation is equal to
the mean of these values (i.e., we ignore any potential skewness in
the error distribution). Each Monte-Carlo realization results in a set
$\calS'_{\rm MW}$, where the prime is used to indicate an element of
the set of $N_{\rm MC}$ Monte Carlo realizations.  Next, for each of
these $\calS'_{\rm MW}$ we compute $\bar{Q}'_{\rm MW}$ and the
corresponding $\calP'_{\rm MW}$ using the same method as described
above (i.e., by comparing $\calS'_{\rm MW}$ to each of the $K=10,000$
model realizations). The resulting distribution of $\calP'_{\rm MW}$
indicates the uncertainty on $\calP_{\rm MW}$ arising from the
uncertainties on the individual $\Vmax$ measurements.

We have performed the above analysis for 8 different host halo masses,
$\log[M_0/(\Msunh)] = 11.0, 11.2, 11.4, ..., 12.4$, two different
cosmologies (`WMAP7' and `Planck'), and 3 different values for the
number of subhaloes in each set, $N=5, 9$ and
$20$. Fig.~\ref{Fig:Qstat} shows the results for the WMAP7 cosmology
and $N=20$, for four different values of $M_0$, as indicated. The
solid, green curve in the upper panels indicates the cumulative
distribution, $P(>\bar{Q})$, obtained from the 10,000 model
realizations. This $P(>\bar{Q})$ is found to be virtually independent
of host halo mass and cosmology, which is a consequence of the fact
that we have normalized $Q$ and that the {\it shape} of the subhalo
$\Vmax$ function is largely invariant. The shaded histograms indicate
the distributions of $\bar{Q}'_{\rm MW}$, while the red, vertical line
indicates $\bar{Q}_{\rm MW}$. Note how $\bar{Q}_{\rm MW}$ shifts to
larger values with increasing host halo mass, and that it is always
located in the tail of the $\bar{Q}$ distribution of the model
predictions, even for halo masses as low as $10^{11.4} h^{-1}\Msun$.

The lower panels of Fig.~\ref{Fig:Qstat} show the cumulative $\Vmax$
distribution, $P(>\Vmax)$, of the MW (red histogram), compared to
the cumulative distribution of the median of the model realizations
(green histogram). The latter is obtained by computing the median
$\Vmax$ of the $i^{\rm th}$ ($i=1,2,...,20$) member of all 10,000
model realizations, and roughly corresponds to the typical $\Vmax$
distribution of the model. The value of the $Q$-statistic
corresponding to these two cumulative distributions is indicated in
each panel, and (as expected) is similar to $\bar{Q}_{\rm MW}$
(indicated by the solid, vertical line in the upper panels). For a MW
host halo mass of $M_0 \sim 10^{12} h^{-1}\Msun$, there is a dramatic
discrepancy between model and data: Not only does the model predict
smaller $\Vmax$ for the Magellanic clouds, it also overpredicts
$\Vmax$ for all other satellites; this is a manifestation of the TBTF
problem. For a MW host halo mass of $M_0 \sim 10^{11.4} h^{-1}\Msun$,
on the other hand, model and data are in remarkably good agreement for
the satellites with $\Vmax < 25\kms$. However, the model now predicts
much lower $\Vmax$ values for the Magellanic clouds (of the order of
$30 - 40 \kms$). There is also some tension at the low $\Vmax$ end
($\Vmax \lta 15\kms$), but this reflects the onset of the `missing
satellites' problem, and may well reflect observational incompleteness
in the inventory of MW satellites. 

The thin, blue histograms show the $P(>\Vmax)$ distributions for the model
realizations that reveal a $\Vmax$-gap similar to the MW (i.e.,
$N(\Vmax>25\kms) = N(\Vmax>55\kms) = 2$). In the case of $M_0 =
10^{11.8}\Msunh$ these are the same 6 models depicted in
Fig.~\ref{Fig:CircularVelocityProfiles}, while for $M_0 =
10^{12.0}\Msunh$ only 1 (out of 10,000) model realization meets these
criteria. For $M_0 = 10^{11.4}\Msun$ ($10^{11.6}\Msunh$) we only show
a random subset of 10 realizations, from a total of 15 (16) out of
10,000 that meet these gap-criteria. The $\bar{Q}$ values
corresponding to these model realizations are indicated as blue,
downward pointing arrows in the upper panels.
\begin{figure}
\centerline{\psfig{figure=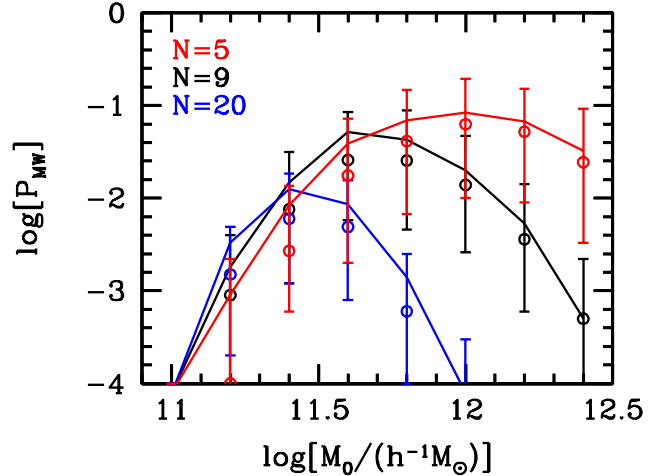,width=\hssize}}
\caption{Solid lines show the host halo mass dependence of the
  probability $\calP_{\rm MW}$ that the WMAP7 cosmology yields host
  haloes with $\bar{Q}$ values as large as that of the MW.  Open
  circles and error-bars indicate the median and the 18 and 84
  percentiles of the corresponding $\calP'_{\rm MW}$ distributions,
  obtained from 10,000 Monte-Carlo realizations of the
  $\Vmax$-distribution of MW satellites. Results are shown for three
  different values of $N$, as indicated.}
\label{Fig:Qprob}
\end{figure}

In the case of $M_0 = 10^{12}\Msunh$, it is clear that TBTF is not
only a problem of an unexpected $\Vmax$-gap between the second and
third ranked members. After all, the model that displays a gap similar
to that in the MW (which in itself is extremely rare), still is an
extremely poor match to the $\Vmax$ distribution of the other
satellite galaxies (at least below $\Vmax \sim 20\kms$)! In particular,
the $\Vmax$ distribution of MW dwarf spheroidals is much broader than
what is predicted by the model. Interestingly, in the case of $M_0 =
10^{11.4}\Msunh$, the models that reveal a MW-like $\Vmax$-gap (which
has an occurrence rate of $\sim 1.5\%$) basically are a good match to
the {\it entire} $\Vmax$ distribution of dwarf spheroidals, at least
down to $\sim 10\kms$. If it wasn't for the fact that such a low halo
mass for the MW is basically ruled out by the timing argument (see e.g.,
Phelps , Nusser \& Desjacques 2013), the space motion of Leo I (e.g.,
Boylan-Kolchin \etal 2013) and the kinematics of blue horizontal
branch stars (e.g., Xue \etal 2008; Kafle \etal 2014) one might be
tempted to consider this strong evidence in support of a MW halo
mass of the order of $10^{11.4} \Msunh$.

Fig.~\ref{Fig:Qprob} summarizes the results of our $Q$-statistic
analysis. The solid lines plot the probability $\calP_{\rm MW}$ as
function of host halo mass for three different values of $N$, as
indicated, while the open circles and error-bars indicate the median
and the 18 and 84 percentiles of the corresponding $\calP'_{\rm MW}$
distributions.  These results are all based on the WMAP7 cosmology,
but the results for the Planck cosmology are virtually
indistinguishable. If we only focus on the five subhaloes with the
largest $\Vmax$ values, then the inference is that the most probable
mass for the MW host halo is $\sim 10^{12} \Msunh$, in good agreement
with a variety of constraints (e.g., Xue \etal 2008; McMillan 2011;
Boylan-Kolchin \etal 2013; Kafle \etal 2014). Furthermore, the
probability that a random $\Lambda$CDM dark matter halo of that mass
has a $\Vmax$ distribution similar to that of the MW (in terms of the
Q-statistic) is $6.3^{+13}_{-5.3}\%$ (68\% CL, when accounting for the
errors on the individual $\Vmax$ measurements of the MW satellites).
This does not signal concern of a potential problem for the
$\Lambda$CDM paradigm. 

However, when using $N=20$ (i.e., when using the $\Vmax$ values of all
MW satellites in Table~1 down to Willman 1, which has $\Vmax
=8.3^{+2.7}_{-0.8}\kms$), a very different picture emerges.  In
particular, the probability that a host halo of $10^{12} \Msunh$ has a
$\Vmax$ distribution similar to that of the MW is now reduced to $<5
\times 10^{-4}$ (at 68\% CL), while the most probable MW mass has
dropped to $M_0 \sim 10^{11.4}\Msunh$ (where 
$\calP_{\rm MW} = 0.6^{+1.2}_{-0.5}\%$ ).  
As is evident from the lower right-hand panel
of Fig.~\ref{Fig:Qstat}, the problem is not just a large gap between
$\sim 25$ and $55\kms$, but rather a problem regarding the {\it
  overall width} of the $\Vmax$ distribution of the MW
satellites. Taking into account that our inventory of MW satellites
may still be incomplete below $\Vmax \sim 15\kms$, a more robust
assessment of TBTF only uses the data of the 9 highest-$\Vmax$
satellite galaxies (LMC, SMC, Sagittarius, Bootes II, Draco, Ursa
Minor, Fornax, Sculptor and Leo I). In that case, we infer a
MW-consistent fraction, for a host halo of $10^{12} \Msunh$, of
$\calP_{\rm MW} = 1.4^{+3.3}_{-1.1}\%$ (68\% CL).
It remains to be seen how the $\calP_{\rm MW}$
  for $N=20$ changes as more and better data for the population of MW
  dSphs continues to come available.


\section{Summary}
 \label{Sec:Summary}
 
In this paper we have used semi-analytical models for the substructure
of dark matter haloes to assess the TBTF problem. We demonstrated that
the model accurately reproduces the average subhalo mass and velocity
functions, as well as their halo-to-halo variance, in high resolution
$N$-body simulations. We then used the model to construct thousands of
realizations of MW-size host haloes, which we used to investigate the
TBTF problem with unprecedented statistical power. 

Previous studies have formulated TBTF in different ways, which we
refer to as the massive subhalo formulation, the gap formulation, and
the density formulation. We have assessed TBTF in all three
formulations, and compared our results to the ELVIS suite of 48
high-resolution zoom-in simulations of MW-sized host haloes
(Garrison-Kimmel \etal 2014a). Overall, the results from our
semi-analytical model are in excellent agreement with the ELVIS
results. Using 100 mock ELVIS suites, we demonstrate, though, that the
suite-to-suite variance is significant, and that a proper assessment
of TBTF statistics requires simulation suites that are an order of
magnitude larger than ELVIS. This motivates the use of semi-analytical
models such as those presented here. 

Regarding the three aforementioned formulations, our assessment of
TBTF is as follows:
\begin{itemize}

\item {\bf massive subhalo formulation:} according to this
  formulation, the MW has a deficit of massive subhaloes, defined as
  subhaloes with $\Vmax > 25\kms$ and $\Vacc > 30\kms$. Whereas the MW
  has only two subhaloes that (most likely) meet these criteria,
  namely the two Magellanic clouds, most MW size host haloes have a
  significantly larger number of `massive subhaloes', $N_{\rm MS}$. We
  find that $\langle N_{\rm MS} \rangle \simeq 7$ (8) and with
  $P(N_{\rm MS}\leq 2) \simeq 0.1\%$ ($0.05\%$) for a host halo mass
  of $M_0 = 10^{12} \Msunh$ in a WMAP7 (Planck) cosmology. However,
  $P(N_{\rm MS}\leq 2)$ has a strong mass dependence, increasing to $>
  10\%$ for $M_0 < 10^{11.8} \Msunh$. In addition, it also depends
  strongly on the exact definition of `massive subhaloes'. For example,
  changing the lower limit on $\Vmax$ from $25\kms$ to $30\kms$ boosts
  $P(N_{\rm MS}\leq 2)$ by an order of magnitude for $M_0 = 10^{12}
  \Msunh$.  Finally, it is important to stress that while lowering the
  mass of the MW's halo increases $P(N_{\rm MS}\leq 2)$, it {\it
    decreases} the probability that such a halo hosts two subhaloes
  comparable to the Magellanic clouds.

\item {\bf gap formulation:} according to this formulation, the MW has
  an unexpectedly large gap in the $\Vmax$-rank-ordered list of its
  satellite galaxies. In particular, no satellite galaxies are known
  with a $\Vmax$ between that of the SMC ($\Vmax = 60 \pm 5 \kms$) and
  that of the Sagittarius dSph ($\Vmax = 25.1 \pm 1.5\kms$). If we
  define `MW-consistent' as having two subhaloes with $\Vmax \geq
  55\kms$ and at most one subhalo with $25 \leq \Vmax < 55\kms$, then
  we find that the MW consistent fraction peaks at $\sim 0.6\%$ around
  a host halo mass of $M_0 \simeq 10^{11.8}\Msunh$. For $M_0 > 10^{12}
  \Msunh$ this fraction is found to be less than $0.1\%$, with little
  dependence on cosmology. As for the massive subhalo count, though,
  these MW consistent fractions are extremely sensitive to the exact
  definition of the gap; for example, changing the $\Vmax$-values of
  the gap by a mere $5\kms$ can change the MW-consistent fraction by
  an order of magnitude.

\item {\bf density formulation:} according to this formulation, the
  densities of the (more massive) subhaloes are too high compared to
  those of the subhaloes hosting MW dSphs. In particular, PZ12
  introduced a density parameter, $\Gamma$, defined by
  Eq.~(\ref{Eq:Gamma}), and argued that whereas all MW dSphs have $0 <
  \Gamma < 1$, the majority of MW-size host haloes have at least one
  subhalo with $\Gamma \geq 1$ (such a subhalo is considered too big,
  or rather dense, to fail). We find that, in a WMAP7 cosmology, only
  $\sim 4.5\%$ ($15.8\%$) of host haloes with $M_0 = 10^{12} \Msunh$
  ($10^{11.8} \Msunh$) are MW-consistent in that $\Gamma_{\rm max} <
  1$. Note that PZ12, using a semi-analytical model similar to ours,
  claimed that these MW-consistent fractions are significantly higher,
  at $10\%$ and $40\%$, respectively. As discussed in the text, we
  believe that their results are hampered by the use of inaccurate
  halo merger trees, and the oversimplified assumption that subhaloes
  have NFW density profiles. Finally, we emphasize that of all three
  TBTF formulations, the density formulation is the one most sensitive
  to cosmology. In particular, we find that, in a Planck cosmology,
  the MW-consistent fraction of host haloes with $M_0 = 10^{12}
  \Msunh$ is only $0.6\%$, rather than $4.5\%$ in a WMAP7 cosmology.

\end{itemize}

An important, and troubling, downside with all three formulations
above is that they suffer from the look-elsewhere effect and/or
disregard certain aspects of the data on the MW satellite population.
In an attempt to remedy these shortcomings, we have devised a new
statistic which is `blind', in that it doesn't require the
pre-selection of some particular scale (i.e., a particular $\Vmax$,
$\Vacc$ and/or $\Gamma$ value), treats all data on equal footing, and
also allows for a straightforward treatment of errors in the $\Vmax$
measurements of individual MW satellites. Similar in spirit to the
Kolmogorov- Smirnov (KS) test, this $Q$-statistic, defined by
Eq.~(\ref{Qstat}), compares two cumulative distributions.  It provides
a (conveniently normalized) measure for the absolute value of the area
between two cumulative distributions, and has the advantage over the
KS-test that it has equal sensitivity throughout the distributions. As
a consequence, it is equally sensitive to the offset between two
distributions as to a difference in the spread.

Using this $Q$-statistic to compare the $\Vmax$ distribution of the 9
MW satellites with the largest $\Vmax$ measurements to that of dark
matter subhaloes in our model realizations, we infer a MW-consistent
fraction, for a host halo of $10^{12} \Msunh$, of $\calP_{\rm MW} =
1.4^{+3.3}_{-1.1}\%$, where the upper and lower bound reflect
uncertainties due to the errors on the individual $\Vmax$ errors. This
is similar to the MW-consistent fraction inferred in the density
formulation, but significantly larger than what is inferred from the
gap-statistic. However, the latter is severely hampered by the
look-elsewhere effect, and does not reflect a proper assessment of the
TBTF severity.

To conclude, it is difficult to express the severity of TBTF in a
single number.  In general, TBTF is (slightly) more problematic in a
Planck cosmology, compared to a WMAP7 cosmology, and is minimized if
the virial mass of the MW's host halo falls in the range $3$ - $6
\times 10^{11} \Msunh$. Such a low MW mass, however, is basically
ruled out by a variety of independent constraints (e.g., Xue \etal
2008; McMillan \etal 2011; Boylan-Kolchin \etal 2013; Phelps \etal
2013; Kafle \etal 2014). Assuming instead a host halo mass of $10^{12}
\Msunh$, and using only data on the 9 known satellite galaxies with
$\Vmax > 15 \kms$, both the density formulation and the $Q$-statistic
suggest a MW-consistent fraction of the order of a few percent,
something which we do not consider particularly challenging for the
$\Lambda$CDM paradigm.  However, if it turns out that the inventory of
MW satellite galaxies is complete to $\sim 8 \kms$ (i.e., future
surveys uncover no new MW satellite galaxies with $\Vmax \gta 8\kms$),
then it is clear that the spread in $\Vmax$ values for the 20 highest
$\Vmax$-ranked satellites is utterly inconsistent with $\Lambda$CDM
predictions. In that case, we either (i) have systematically
underestimated the $\Vmax$ values (by about a factor of two) of all MW
dwarf spheroidals, (ii) the process of galaxy formation significantly
lowers the central densities (and hence $\Vmax$) of the subhaloes
hosting dwarf spheroidals, or (iii) the $\Lambda$CDM paradigm is
actually falsified. In this respect, it remains to be seen whether
baryonic effects, such as those discussed in Zolotov \etal (2012),
Brooks \& Zolotov (2014), and Arraki \etal (2014), can give rise to
satellite populations in hydro-dynamical simulations with $\Vmax$
distributions in better agreement with observations. We hope that the
$Q$-statistic introduced here will prove useful to compare such
simulations to the data.


\section*{Acknowledgments}

We are grateful to the people responsible for the Bolshoi and ELVIS
simulations for making their halo catalogs publicly available, the
anonymous referee for an insightful report that helped to improve the
presentation, and to the following individuals for their advice and
useful discussions: Peter Behroozi, Mike Boylan-Kolchin, Shea
Garrison-Kimmel, Andrew Hearin, and Duncan Campbell.



\label{lastpage}

\end{document}